\documentclass{emulateapj}

\usepackage{amsmath} 
\usepackage{amsfonts} 
\usepackage{amssymb} 
\usepackage{xspace}
\usepackage{wasysym}
\usepackage{mdwlist}
\newcommand{\kepler}{$Kepler$\xspace}

\newcommand{\msun}{\ensuremath{M_\odot}\xspace}		     
 		     
\newcommand{\rsun}{\ensuremath{R_\odot}\xspace}

\newcommand{\mearth}{\ensuremath{M_{\oplus}}\xspace}

\newcommand{\rearth}{\ensuremath{R_{\oplus}}\xspace}

\newcommand{\insol}{\ensuremath{S_{\oplus}}\xspace} 	

\newcommand{\kic}{KIC~8120608\xspace}
\newcommand{\starname}{Kepler-186\xspace}
\newcommand{\planetb}{Kepler-186b\xspace}
\newcommand{\planetc}{Kepler-186c\xspace}
\newcommand{\planetd}{Kepler-186d\xspace}
\newcommand{\planete}{Kepler-186e\xspace}
\newcommand{\planetf}{Kepler-186f\xspace}

\def\deg{^\circ}

\usepackage[normalem]{ulem}
\usepackage{soul}
\usepackage{color}
\definecolor{blue}{RGB}{0,0,255}
\definecolor{orange}{RGB}{255,99,71}

\def\aa{Astron. Astrophys.}

\def\aap{Astron. Astrophys.}

\def\aj{Astron. J.}

\def\apj{Astrophys. J.}

\def\apjl{Astrophys. J. Letters}

\def\apjs{Astrophys. J. Suppl.}

\def\araa{Ann. Rev. Astron. Astrophys.}

\def\astrb{Astrobiology}

\def\icarus{Icarus}

\def\jas{J. Atmosph. Sciences}

\def\jgr{J. Geophys. Res.}

\def\mnras{Monthly Not. Royal Astron. Soc.}

\def\nat{Nature}

\def\planss{Planet. Space Science}

\shortauthors{Bolmont et al.}
\shorttitle{Formation, tidal evolution and habitability of the Kepler-186 system}
\begin{document}
\bibliographystyle{apj}
\title{Formation, Tidal Evolution and Habitability of the Kepler-186 System}
\author{
Emeline~Bolmont\altaffilmark{1,2,*}, 
Sean~N.~Raymond\altaffilmark{1,2},
Philip~von Paris\altaffilmark{3,+},
Franck~Selsis\altaffilmark{1,2},
Franck~Hersant\altaffilmark{1,2},
Elisa~V.~Quintana\altaffilmark{4,5} and
Thomas~Barclay\altaffilmark{5,6} 
}
\altaffiltext{1}{Univ. Bordeaux, Laboratoire d'Astrophysique de Bordeaux, UMR 5804, F-33270 Floirac, France}
\altaffiltext{2}{CNRS, Laboratoire d'Astrophysique de Bordeaux, UMR 5804, F-33270 Floirac, France}
\altaffiltext{3}{Institut f\"{u}r Planetenforschung, Deutsches Zentrum f\"{u}r Luft- und Raumfahrt (DLR), Rutherfordstr. 2, D-12489 Berlin, Germany}
\altaffiltext{4}{SETI Institute, 189 Bernardo Ave, Suite 100, Mountain View, CA 94043, USA}
\altaffiltext{5}{NASA Ames Research Center, Moffett Field, CA 94035, USA}
\altaffiltext{6}{Bay Area Environmental Research Institute/NASA Ames Research Center, Moffett Field, CA 94035, USA}

\altaffiltext{+}{Current address: Univ. Bordeaux, Laboratoire d'Astrophysique de Bordeaux, UMR 5804, F-33270 Floirac, France and CNRS, Laboratoire d'Astrophysique de Bordeaux, UMR 5804, F-33270 Floirac, France}

\altaffiltext{*}{Author to whom all correspondence should be addressed. email: bolmont@obs.u-bordeaux1.fr}

\begin{abstract}
The \starname system consists of five planets orbiting an early M dwarf. The planets have physical radii of 1.0--1.50 \rearth and orbital periods of 4--130 days. The $1.1 \rearth$ \planetf with a period of 130\,days is of particular interest. Its insolation of roughly $0.32$ \insol places it within the surface liquid water habitable zone (HZ). We present a multifaceted study of the \starname system, using two sets of parameters which are consistent with the data and also self-consistent. First, we show that the distribution of planet masses can be roughly reproduced if the planets were accreted from a high surface density disk presumably sculpted by an earlier phase of migration. However, our simulations predict the existence of one to two undetected planets between planets e and f. Next, we present a dynamical analysis of the system including the effect of tides. The timescale for tidal evolution is short enough that the four inner planets must have small obliquities and near-synchronous rotation rates. The tidal evolution of \planetf is slow enough that its current spin state depends on a combination of its initial spin state, its dissipation rate, and the stellar age. Finally, we study the habitability of \planetf with a one-dimensional climate model. The planet's surface temperature can be raised above 273 K with 0.5--5 bars of $\mathrm{CO_2}$, depending on the amount of $\mathrm{N_2}$ present. \planetf represents a case study of an Earth-sized planet in the cooler regions of the HZ of a cool star.
\end{abstract}

\keywords{methods: numerical --  planets and satellites: atmospheres -- planets and satellites: dynamical evolution and stability -- planets and satellites: formation -- stars: individual (\starname, \kic)}


\section{Introduction}

The \kepler mission~\citep{borucki2010} has made key discoveries on the road to finding Earth-like planets \citep[e.g.,][]{Batalha.etal.:2011,Borucki.etal:2012,Fressin.etal.2012,Borucki.etal:2013}. The recent detection of an Earth-sized planet in the HZ of an M star (i.e., the \starname system, \citealp{Quintana2014}) brings us a step closer to finding a true Earth twin. 

\begin{figure}[htbp]
\includegraphics[width=0.45\textwidth]{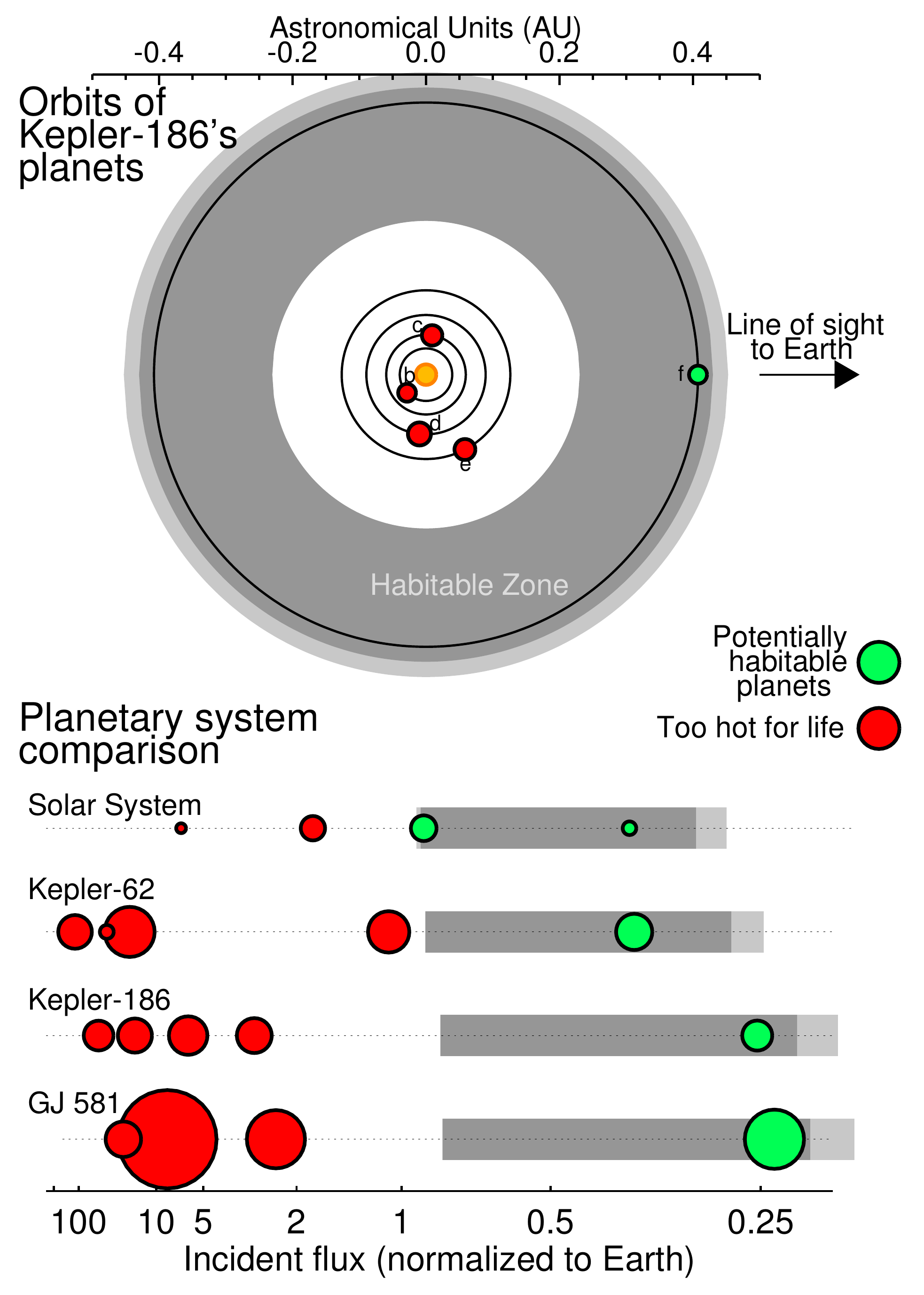}
\caption{Orbital configuration of the \starname system. The top part shows a top-down view of the system, assuming orbits from set $\mathcal{A}$.  The habitable zone boundaries are from \cite{kopparapu2013}: the inner boundaries are the moist/runaway greenhouse limits and the outer boundaries are the maximum greenhouse and early Mars limits. The sizes of the symbols are not to scale with the planetary orbits. The bottom part of the plot shows a comparison between four different planetary systems which contain planets in the HZ: the solar system, Kepler-62~\citep{Borucki.etal:2013}, Kepler-186~\citep{Quintana2014}, and GJ 581~\citep{udry07,mayor09}. Note that the inner moist and runaway greenhouse limits of the habitable zone are the same for Kepler-62, Kepler-186, and GJ~581. Given the consistent insolation scaling, the x axis is linear in orbital distance but the scale is different for each system.  The planets' relative sizes are correct, although for GJ~581 the planetary radii were calculated as $R = [M sin(i)]^{2.06}$, following \cite{Lissauer.etal:2011}.}
\label{Kepler186_HZ}
\end{figure}

The \starname planetary system hosts five known planets including \planetf, an Earth-sized planet in the HZ \citep{Selsis2007,kopparapu2013}. Figure \ref{Kepler186_HZ} shows a comparison between the \starname system, the solar system, and two other systems with potentially habitable planets: Kepler-62~\citep{Borucki.etal:2013} and GJ~581~\citep{udry07,mayor09}. Climate models have shown that GJ~581d, a super-Earth near the outer edge of the HZ of its host M star, could sustain surface liquid water \citep[e.g.,][]{wordsworth2010}. \planetf receives a comparable or perhaps slightly higher stellar flux than GJ~581, placing it more comfortably within the HZ.

Here, we use the definition of the classical HZ \citep[HZ, e.g., ][]{dole1964,hart1978,kasting1993,Selsis2007,kopparapu2013}. Acknowledging the fact that all terrestrial life needs liquid water, the HZ is defined as the region around a star where a terrestrial planet can host liquid water on the surface. The extent of this HZ naturally depends on the atmospheric conditions (composition, pressure) as well as on the properties of the central star. Many more factors influence the width of the HZ, such as the geological activity \citep[e.g.,][]{lammer2010}, the biosphere itself \citep[e.g.,][]{grenfell2010}, or the dynamical environment of the planetary system \citep[e.g.,][]{menou03,barnes04,jones2006,sandor2007,kopparapu2010}.

\medskip

We present a three-pronged study of the \starname system.  We first try to reproduce the orbital architecture of the system using simple accretion simulations (Section \ref{formation}).  We show that certain features of the system--such as the large gap between planets e and f--are hard to explain. We next briefly discuss the long-term dynamical stability of the system (Section \ref{stability}). In Section \ref{tidal}, we study the long-term dynamical, tidal, and spin evolution of the system. We use both simple tidal models and $N$-body simulations which include both tides and general relativity.  Next, we study the atmospheric conditions needed to bring \planetf's surface temperature into the liquid water range (Section \ref{habi}). We discuss our findings and conclude in Section \ref{conclusion}.

\newpage
\section{Model Input Parameter}
\label{input}
The stellar properties and planetary parameter given in \citet{Quintana2014} are the median values of each corresponding probability density. However, the median values are not intended to be self consistent (for example, the relationship between the density, radius, and mass of the star is not respected). In order to study the dynamical evolution of the system and its habitability, we need a set of consistent parameters. There are two ways of obtaining a consistent set of values for stellar and planetary parameters. They define what we call set $\mathcal{A}$ and set $\mathcal{B}$: 
\begin{itemize*}
\item[1.] Set $\mathcal{A}$: the stellar properties are chosen to match the point estimate transit model obtained by performing Markov Chain Monte Carlo (MCMC) realizations \citep[e.g.,][]{Barclay2013a};
\item[2.] Set $\mathcal{B}$: the transit model is chosen to match the point estimate stellar properties obtained by performing MCMC realizations. The stellar properties of this set correspond to those of table S1 in \citet{Quintana2014}.
\end{itemize*}

Set $\mathcal{A}$ and $\mathcal{B}$ are both valid sets of parameters, meaning that they are consistent with the data and are also self-consistent.

The stellar mass is $0.5359$~$M_\odot$ in set $\mathcal{A}$ and $0.478$~$M_\odot$ in set $\mathcal{B}$ (Table \ref{tab:star} and \citealt{Quintana2014}). This produces different values for the planets' semi-major axes and insolations  (Table \ref{tab:planet}). The same is true for their planetary radii, as these are derived from the detected transit depth and the stellar radius. The fact that there is such a difference between the parameters values of each set illustrate the limited knowledge we have of the system. We expect the real parameters of the system to be within the range of sets $\mathcal{A}$ and $\mathcal{B}$, so we performed dynamical simulations for both these sets. However, averaging between set $\mathcal{A}$ and $\mathcal{B}$ is not an option in so far as it would not correspond to anything realistic, the self-consistency would be lost \citep[as in][]{Quintana2014}. That is why we chose a different approach than in \citet{Quintana2014} where semi-major axes are not chosen to match the derived stellar properties, but are rather determined by the transit model only.

All five planets in the \starname system have radii between 1.0 and 1.5~$R_\oplus$ for both parameter sets. Given that low-density, gas-dominated planets tend to be larger than $1.5$--$2\rearth$~\citep{Weiss2013,Weiss2014,lopez13,jontofhutter14,Marcy2014,marcy2014b}, all five \starname planets are probably rocky or at least solid. The planets' masses have not been constrained with radial velocity or transit timing measurements \citep{Quintana2014}. Table \ref{tab:plcomp} shows the range of plausible planetary masses assuming a range of compositions: 100\% ice, 50\% ice/ 50\% rock, Earth-like composition, and 100\% iron \citep[following][]{Fortney.etal:2007}. 

\begin{table*}[htbp]
\begin{center}
\caption{Stellar Properties}
\vspace{0.1cm}
\begin{tabular}{ccccc}
\hline     
 & Mass (\msun) & Radius (\rsun) & $T_{\textrm{eff}}$ (K) & $L_\star/L_\odot$ \\
\hline
Set $\mathcal{A}$ & 0.536 & 0.5138 & 3747 & 0.0468 \\
Set $\mathcal{B}$\footnote{From Table S1 of \citet{Quintana2014}} & 0.478 &  0.4720 & 3788 & 0.0412 \\
\hline
\end{tabular} 
\label{tab:star} 
\end{center}
\end{table*}

\begin{table*}[htbp]
\begin{center}
\caption{Planetary Physical Parameters}
\vspace{0.1cm}
\begin{tabular}{ccccccc}
\hline
& & \planetb & \planetc & \planetd & \planete & \planetf \\ 
\hline
Period (days)\footnote{From Table S2 of \citet{Quintana2014}, rounded}  & & 3.887 & 7.267 & 13.34 & 22.41 & 129.9 \\
Semimajor axis (AU) &Set $\mathcal{A}$ & 0.0393 & 0.0596 & 0.0894 & 0.1264 & 0.4078 \\
&Set $\mathcal{B}$ & 0.0378 & 0.0574 & 0.0861 & 0.1216 & 0.3926 \\
Radius (\rearth) &  Set $\mathcal{A}$ & 1.16 & 1.33 & 1.50 & 1.36 & 1.17 \\
&Set $\mathcal{B}$ & 1.08 & 1.25 & 1.39 & 1.33 & 1.13 \\
Insolation ($S_\oplus$) &  Set $\mathcal{A}$ & 30.2 & 13.1 & 5.8 & 2.9 & 0.28 \\
& Set $\mathcal{B}$ & 28.7 & 12.5 & 5.5 & 2.8 & 0.27 \\
\hline
\end{tabular} 
\label{tab:planet} 
\end{center}
\end{table*}

\begin{table*}[htbp]
\begin{center}
\caption{Range of Plausible Planet Masses (in \mearth) Using Formulae from \citet{Fortney.etal:2007} for Set $\mathcal{B}$}
\vspace{0.1cm}
\begin{tabular}{l l l l l}
\hline
& Pure Ice & 50\% Ice-rock & Earth-like & Pure Iron\\
\hline
Planet b & 0.29 & 0.48 & 1.32 & 3.36 \\
Planet c & 0.46 & 0.77 & 2.27 & 6.30 \\
Planet d & 0.65 & 1.10 & 3.45 & 10.2 \\
Planet e & 0.56 & 0.95 & 2.89 & 8.32 \\
Planet f & 0.33 & 0.55 & 1.55 & 4.06 \\
\hline
\end{tabular} 
\label{tab:plcomp}
\end{center}
\end{table*}

\section{Formation}
\label{formation}

\subsection{Formation Models}

At least six candidate mechanisms have been proposed to explain the origin of close-in low-mass planets \citep{raymond08,raymond14pp6}. In theory, these mechanisms can be distinguished using two observable quantities: the inner planetary system architecture and the mean planet density. Given current constraints, the two leading candidates are the collisional growth of a population of inward-migrating planetary embryos~\citep{Terquem.etal:2007,cossou13b} and in situ accretion of a population of close-in planetary embryos \citep{raymond08,Chiang.etal:2013}. These two mechanisms might work in tandem, with an early phase of inward migration followed by a later phase of collisions~\citep{Hansen.etal:2012,raymond14pp6}. 

A problem with the in situ accretion model is that it requires very massive disks close to their stars. For the observed systems of hot Super Earths to have accreted locally, the typical inner disk must be far more massive than that suggested by sub-millimeter observations of outer disks~\citep{raymond08}. In addition, the inner disk must follow a steeper surface profile. Whereas outer disks are observed to follow $r^{-(0.5~{\rm to}~1)}$ radial surface density profiles \citep{Mundy.etal:2000,Andrews.etal:2007a}, a ``minimum-mass extrasolar nebula'' would need to follow an $r^{-1.6}$ profile~\citep{Chiang.etal:2013}. In fact, the minimum-mass disks inferred from the observed systems of hot Super Earths span a wide range of disk profiles, from $r^{0.5}$ to $r^{-3}$~\citep{Raymond:2014}. This is in conflict with accepted disk theory.  Minimum-mass disks created from the distribution of hot super-Earths therefore do not reflect the properties of the nascent (gaseous) protoplanetary disks.  Rather, minimum-mass disks likely represent the distribution of solids in the inner parts of disks after a phase of migration~\citep{Raymond:2014}.

\subsection{Minimum-mass Disk Analysis}

We performed a simple minimum-mass experiment on this system. We first calculated mass estimates for the planets.  Given that they are all smaller than $1.5 \rearth$, we expect the planets to be rocky~\citep{Weiss2013,Weiss2014,lopez13,jontofhutter14,Marcy2014,marcy2014b}.  As a baseline, we assumed that their bulk compositions are the same as Earth's and used a corresponding mass--radius relation\citep[following][]{Valencia.etal:2006} whereby $M \propto R^{3.7}$. We also tested other observationally derived mass--radius relations: $M \propto R^{2.06}$~\citep{Lissauer.etal:2011} and the empirical, flux-dependent relation from \cite{Weiss2013}. To obtain surface densities, we spread the planets' masses into concentric annuli.  We chose the boundaries between adjacent annuli to be the geometric means between the two planets' orbital radii.  For the innermost (outermost) planets we chose the inner (outer) edge by assuming the same spacing as for the next-farthest out (next closest-in) pair of planets.  We then simply fit a power law to the distribution of derived surface densities.  We previously validated this method in~\cite{Raymond:2014}.

\begin{figure}[htbp]
\includegraphics[width=0.45\textwidth]{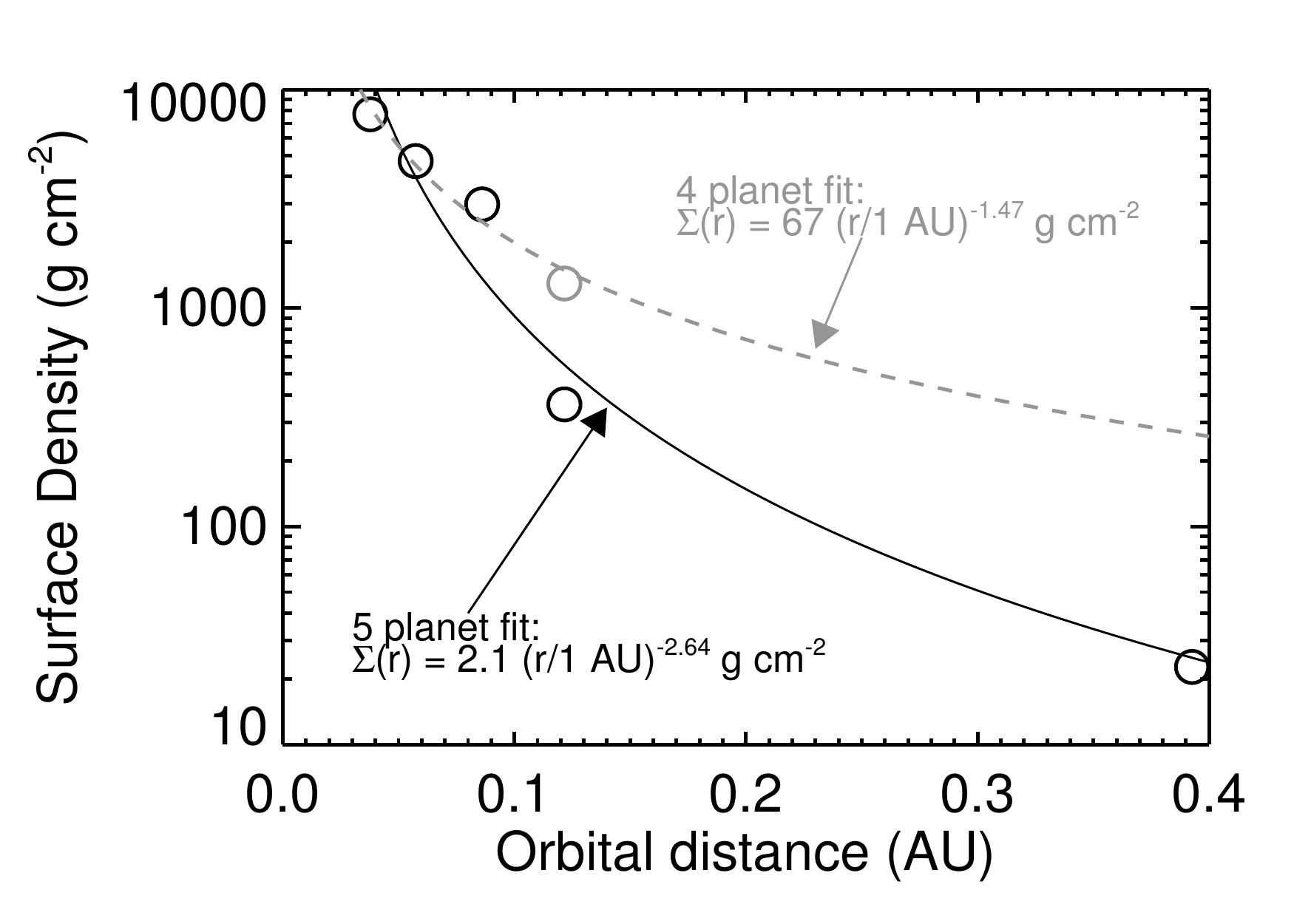}
\caption{``Minimum-mass solar nebula''-type fit to the \starname system. The symbols were calculated assuming an Earth-like composition for all of the planets. The solid curve is the fiducial best fit. The gray dashed curve is the best fit using only the four inner planets.  Note the higher surface density for planet e (at 0.12 AU) in the four planet fit. Although this plot shows fits for set $\mathcal{B}$, the inferred minimum-mass disks were almost identical using set $\mathcal{A}$. }
\label{fig:mmsn}
\end{figure}

Figure~\ref{fig:mmsn} shows the outcome of this experiment. The best-fit, minimum-mass disk has a steep profile: $\Sigma(r) \propto r^{-2.64}$. Using different mass--radius relations, the slope of the surface density profile remained in a well-confined range: $\Sigma(r) \propto r^{-(2.5~{\rm to}~2.7)}$.  Systematic differences in planetary composition correlated with orbital distance could in principle change the slope of this fit, i.e., if more distant planets were preferentially high- or low-density.  However, any strong effect has already been accounted for in the flux-dependent relation from \cite{Weiss2013}, which remained very close to the best-fit value.  We repeated this experiment with both sets $\mathcal{A}$ and $\mathcal{B}$ and found slopes that were the same to within $\pm 0.01$.  The rest of the analysis in this section uses set $\mathcal{B}$ as a comparison sample.

While the four inner planets are in a ``packed'' orbital configuration, there is a large gap between planets $e$ and $f$. It is possible that an additional, as-yet undetected planet could exist within this gap (see below). A minimum-mass disk calculated with all five planets' orbits may therefore be missing material between planets e and f.  We therefore repeated the minimum-mass disk experiment using just the four inner planets.  This will generate a narrower disk but perhaps a more representative one.  By removing the outermost planet (planet f) from the fit, planet e's effective surface density is strongly increased.  This is because, since planet e is much closer to planet d than to planet f, the annulus over which planet e's mass was spread became much narrower.  With a much higher outer surface density, the four planet fit therefore produced a flatter profile with $\Sigma(r) \propto r^{-1.47}$. Although this profile was built using just the inner planets, if we extend the disk to larger orbital radii we inherently assume that an additional planet exists between planets e and f.  This is because there is far more mass in the outer regions than included in planet f alone. If an additional planet does indeed exist, then the slope of the underlying minimum-mass disk would indeed be flatter than the $\Sigma(r) \propto r^{-2.6}$ profile derived above.  Of course, the exact slope would depend on the properties of the additional planet.

The surface density profiles from Figure~\ref{fig:mmsn} can be interpreted as the initial conditions for the final stage of planetary formation. This is {\em after} a phase of inward migration of solids. Viscous disk models predict $r^{-(0.5~{\rm to}~1)}$ profiles. Sub-millimeter observations of the outer parts of protoplanetary disks consistently find the same slopes~\citep[$\Sigma \propto r^{-(0.5~{\rm to}~1)}$; see][and references therein]{Williams.etal:2011}.The very steep profile inferred from our minimum-mass disk analysis ($\Sigma(r) \propto r^{-2.6}$) probably does not represent the state of the gaseous protoplanetary disk.  Rather, this profile represents the distribution of solids in the inner parts of the disk immediately before the final assembly phase.  A phase of inward migration likely shaped this distribution.  This migration could potentially occur when objects are boulder-sized or smaller~\citep{weidenschilling77,boley13,chatterjee14} or when they are approximately Mars-sized or larger~\citep{goldreich80,ward97}. Migration tends to produce systems of planets in chains of mean motion resonances~\citep{Terquem.etal:2007,Ogihara.etal:2009,cossou13b,pierens13}.  This is of course not observed in the Kepler-186 system.  However, later dynamical evolution can extract the planets from resonance~\citep{Terquem.etal:2007,cossou13b}.  In fact, this type of evolution may be widespread.  A late stage of giant impacts is naturally triggered when the gaseous disk disperses due to the disappearance of the associated damping forces~\citep{cossou13b}.

\subsection{Accretion Simulations}

We attempted to reproduce the \starname system numerically.  We started from the minimum-mass disks in Figure~\ref{fig:mmsn} which were presumably already sculpted by migration. We performed two sets of $N$-body simulations of late-stage accretion of planetary embryos and planetesimals, in disks with surface density profiles $\Sigma \propto r^{-2.64}$ and $\Sigma \propto^{-1.47}$. Each set consisted of 10 simulations.  Our initial conditions consisted of populations of 40 planetary embryos and 400 planetesimals spread between 0.03 and 0.5 AU (the approximate radial extent of the Kepler-186 planets).  The final stages of accretion are stochastic: slightly different initial conditions yield different outcomes.  This is true even with infinite resolution because the stochastic nature of this process comes from individual scattering events between embryos.  At late times, there is invariably a small number of embryos such that ``shot noise'' produces stochastic behavior.  Our chosen resolution is a compromise between adequately capturing physical effects such as dynamical friction~\citep[e.g.][]{raymond06,obrien06} and computational expense.  Simulations with comparable resolution have indeed been shown to capture the key aspects of accretion~\citep{raymond05b,Kokubo.etal:2006}.  We neglected the outer parts of the planetary system beyond 0.5 AU.  Giant planets in the outer regions can indeed affect the dynamics of the inner regions (e.g., via secular resonances).  However, given the absence of constraints we prefer to keep our setup as simple as possible.

The total initial mass in planetary embryos and planetesimals was $15 M_\oplus$.  This is slightly more than the minimum-mass disks but is needed to produce planets comparable to the observed ones.  Early simulations with lower-mass disks consistently underestimated the planets' masses. Embryos were given physical densities of 3 $\mathrm{g~cm^{-3}}$. This is comparable to the densities of the Moon and Mars.  The embryos' initial inclinations were randomly chosen between zero and $0.1$~degree. Each system was integrated with the {\tt Mercury} hybrid integrator~\citep{Chambers:1999} for 10 Myr using a 0.2 day timestep. Collisions were treated as perfect mergers and gas effects were not included. Given the short lifetimes of gaseous protoplanetary disks~\citep{Haisch.etal:2001,Hillenbrand:2008}, the assumption of a gas-free environment may not be realistic. However, this could be justified if we assume that our initial conditions were sculpted by an earlier phase of migration, and thus represent the state of the disk just after the dissipation of the gaseous disk. In that case, our choices of embryo and planetesimal masses have little effect on the outcome~\citep{Kokubo.etal:2006}. 

\begin{figure}[htbp]
\includegraphics[width=0.45\textwidth]{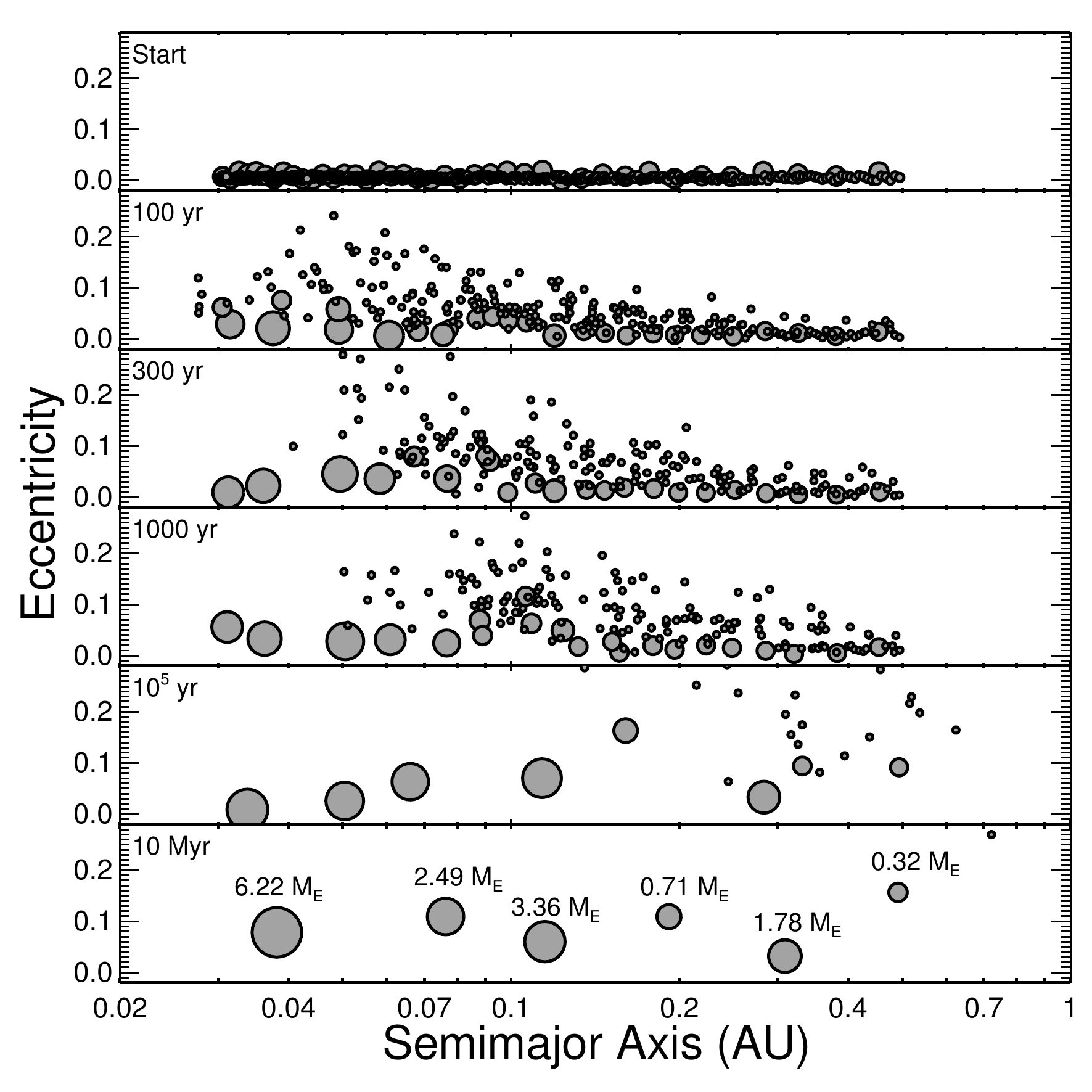}
\caption{Evolution of a simulation of terrestrial accretion. Each circle corresponds to a growing planetary embryo or planetesimal with radius $R \propto M^{0.27}$~\citep[][not to scale on the $x$ axis]{Valencia.etal:2006}. The masses of the final planets are labeled in the last panel. }
\label{fig:aet}
\end{figure}

Figure~\ref{fig:aet} shows the evolution of a simulation in an $r^{-2.64}$ disk that formed six planets interior to 0.5 AU. Accretion was fast and proceeded as a wave sweeping outward. The outward sweeping is caused by the radial dependence of the eccentricity excitation and encounter timescales. Accretion was mostly finished interior to 0.1 AU within $10^5$ yr and is complete throughout the disk by 1-10 Myr. This is characteristic of all of our simulations, although the simulations in flatter disks ($\Sigma \propto r^{-1.47}$) were slower than in steeper disks~\citep[$\Sigma \propto r^{-2.64}$; as expected][]{raymond05b}.  In the simulation from Figure~\ref{fig:aet}, three planets formed between 0.03 and 0.11 AU, the range occupied by four planets in the Kepler-186 system.  The simulation produced three additional planets including a reasonable analog to Kepler-186f at 0.31 AU.  One of the extra planets was located at 0.19 AU, in the empty gap between planets $e$ and $f$.

\begin{figure}[htbp]
\includegraphics[width=0.45\textwidth]{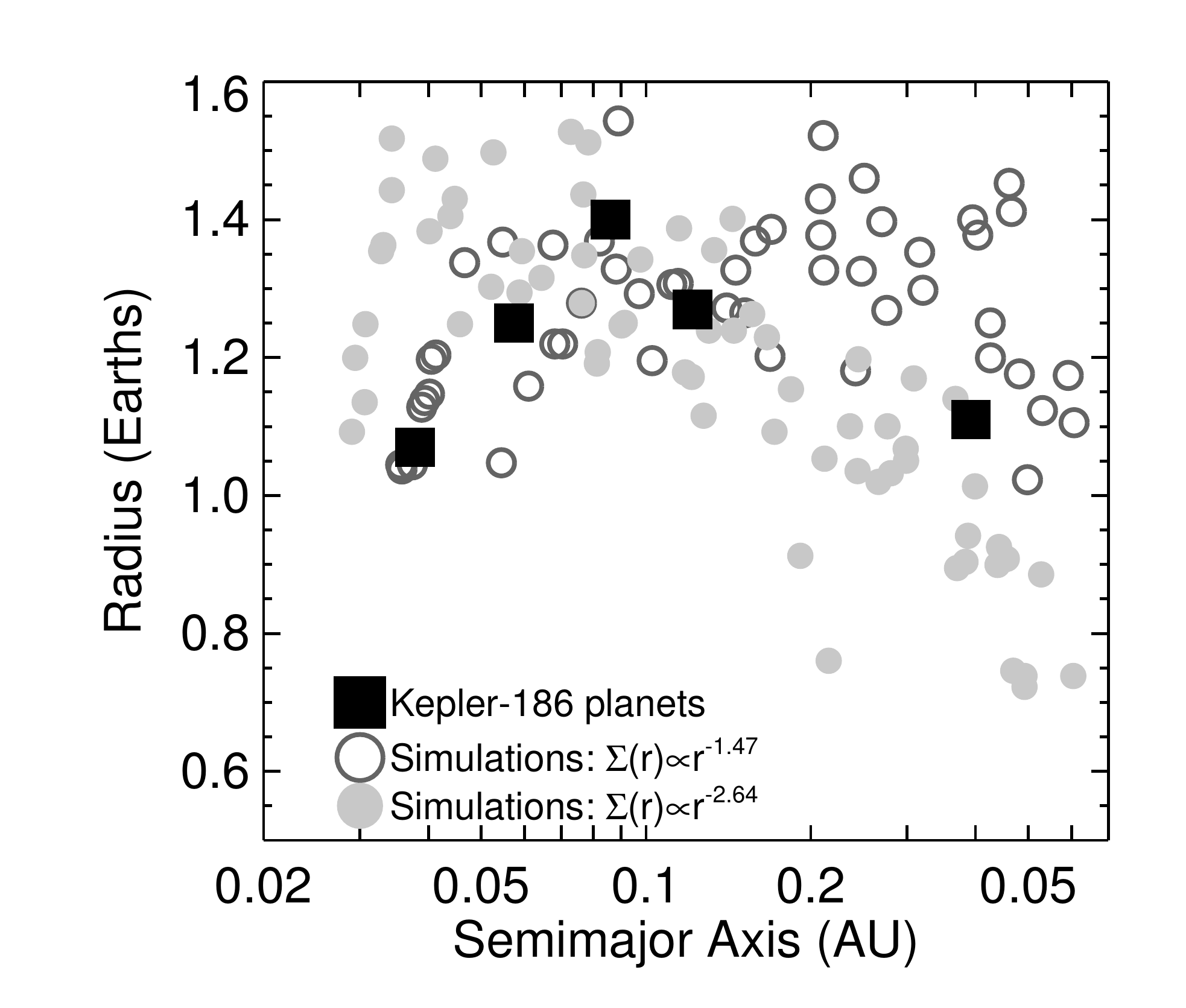}
\caption{Size vs orbital radius for the \starname system (large black squares) compared with different suites of $N$-body simulations in disks with different properties (symbols).  }
\label{fig:m-a}
\end{figure}

Figure~\ref{fig:m-a} shows the mass versus orbital radius distribution for the $N$-body simulations, compared with the actual Kepler-186 system. The radii of the simulated planets were calculated assuming an Earth-like composition ($R \propto M^{0.27}$~\citep{Valencia.etal:2006}). The distributions of accreted planets clearly retain a ``memory'' of their initial disk profiles~\citep{raymond05b}. The planets that formed in disks with a steeper ($r^{-2.64}$) surface density profile were more massive close in and smaller farther out compared with the planets formed within the shallower ($r^{-1.47}$) profile. 

The disk with a shallower ($r^{-1.47}$) profile provides a better fit to the inner parts of the system's mass distribution.  However, neither set of simulations does a good job of fitting the outer parts.  Planet f's size is systematically underestimated in simulations with a steeper ($r^{-2.64}$) disk and systematically overestimated in simulations with a flatter ($r^{-1.47}$) disk. The two innermost planets' sizes are also systematically overestimated in the steeper disk.  Given that the flatter disk profile was built using just the four inner planets, it is reassuring that the planets which formed in such a disk do indeed roughly match the mass distribution within $\sim 0.1$~AU.  The fact that more distant planets are more massive than the real planet f is not surprising.

The inclinations of the simulated planets were too large to be consistent with the true system.  Of particular importance is the mutual inclination between two planets' orbits $\Phi_{12}$, calculated as:
\begin{equation}
\cos{\Phi_{12}} = \cos{i_1}\cos{i_2} + \sin{i_1}\sin{i_2}cos{(\Omega_1-\Omega_2)},
\end{equation}
where $i$ denotes each planet's inclination and $\Omega$ is the longitude of the ascending node. 

We have no information about the mutual inclinations between the planets' orbits in the system.  Such a measurement can only be made during a special event such as a planet-planet eclipse~\citep{ragozzine10,hirano12}.\footnote{A planet--planet eclipse may in fact have occurred in the Kepler-186 system on June 15th, 2014.  See this link: http://planetplanet.net/2014/06/04/something-amazing-will-happen-on-june-15th-but-no-one-on-earth-will-see-it/.}

We can constrain the mutual inclinations of the Kepler-186 planets using statistical arguments.  We know that each planet transits the star.  Therefore, at a given orbital phase (i.e., orbital longitude), each planet's orbit approaches a common plane.  If the planets have large mutual inclinations then the only way for their orbits to line up like this is if both a) the planets are in an inclination-type resonance, meaning that they have similar longitudes of ascending node; and b) our line of sight is aligned (or anti-aligned) with that longitude.  The orbital periods of adjacent planets do not suggest the presence of any resonances in the system.  It therefore seems unlikely that the planets' orbits should be aligned, and therefore unlikely that the planets' orbits have large mutual inclinations.

If we assume that the planets' orbital alignments are uncorrelated, then we can therefore simply constrain each planet's inclination with respect to a common plane.  For simplicity, we assume the common plane to be perfectly aligned with \textit{Kepler's} line of sight.  Each planet's orbit must be inclined by less than a given angle with respect to this plane to remain in transit.  Of course, the critical angle depends on the viewing angle because any orbit will cross a given plane.  In practice, the relevant critical angle is simply the star's angular radius: a planet will necessarily transit if its inclination with respect to a common plane remains less than the angular radius of the star.

\begin{figure}[htbp]
\includegraphics[width=0.45\textwidth]{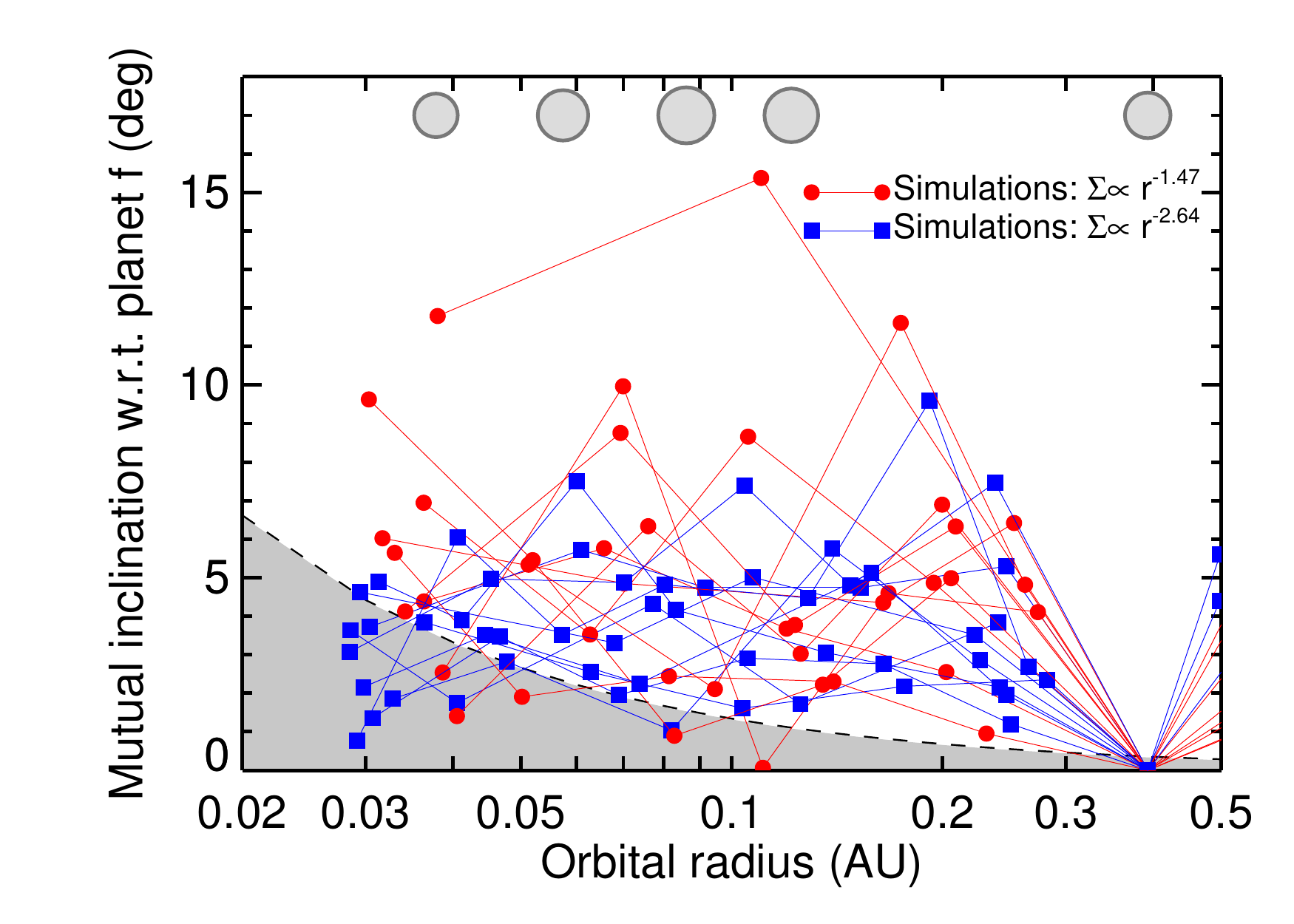}
\caption{Mutual inclinations relative to analogs of planet f in the simulated systems. Orbital radii were scaled such that the planet f analogs were located at the correct distance. The planets' actual positions (set $\mathcal{B}$) are shown by the gray symbols at the top of the plot.}
\label{fig:mut_inc}
\end{figure}

Figure~\ref{fig:mut_inc} shows the mutual inclinations of the systems relative to the closest analog of planet f in each simulation. Planet f analogs were simply defined as the planets closest to the true planet f's orbital radius. The shaded area of Figure~\ref{fig:mut_inc} shows the region where a planet's inclination relative to the plane of planet f's orbit is smaller than the angular size of the star. Since planet f's orbit is used to define a common plane, if its orbit transits then so too does the orbit of any planet in the shaded region.  Above the shaded region a planet is statistically unlikely to transit, unless its longitude of ascending node is close to being aligned with the viewing angle.

It is clear from Figure~\ref{fig:mut_inc} that the simulations produce systems that are too dynamically hot. The mutual inclinations between planets are too large for five planets to be observed in transit~\citep{Raymond:2014}. The observed multiple-planet systems are inferred to typically have mutual inclinations of not more than a few degrees~\citep{Fang2012,Tremaine2012}. These overly large mutual inclinations appear to be an additional strike against the in situ formation mechanism for hot super-Earths.  Additional dissipation is needed to bring the planets' orbits back toward a common plane.  In the context of the inward migration model, the gaseous disk can provide this dissipation, even if the disk's thinning triggers late instabilities~\citep{cossou13b}.  However, we acknowledge that there exist additional mechanisms that can also damp planets' inclinations.  For instance, over gigayear timescales, tidal damping may act to decrease mutual inclinations~\citep[see][]{Hansen2013}.

\subsection{A Missing Planet Between Planets $e$ and $f$?}

\begin{figure}[htbp]
\includegraphics[width=0.45\textwidth]{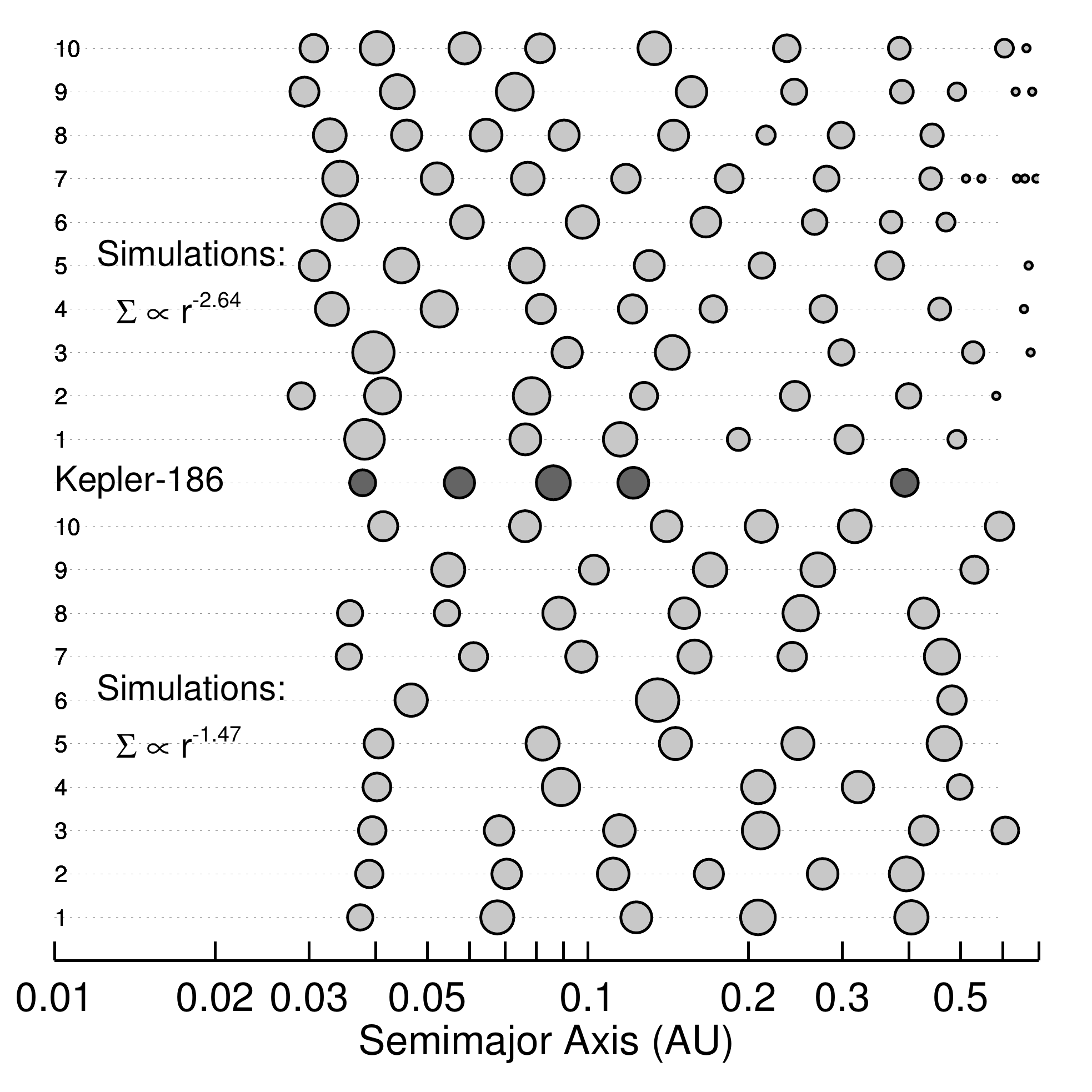}
\caption{Final state of the 20 simulated systems compared with the actual one. The size of each simulated planet was scaled assuming a rocky composition. }
\label{fig:sims_all}
\end{figure}

The most glaring inconsistency between the simulations and the Kepler-186 system is that the simulations form too many planets (Figure~\ref{fig:sims_all}. Our simulations produced 3--8 planets interior to 0.5 AU. Most (12 out of 20) simulations produced 6 or more planets. In all simulations, at least one planet formed between the orbits of known planets e and f, in the range 0.15--0.4 AU. In the simulations that formed five planets, the inner parts of the systems tended to have less tightly packed orbital configurations than the real one. However, the outer parts of these systems (from 0.1--0.4 AU) were more packed.

\begin{figure}[htbp]
\includegraphics[width=0.45\textwidth]{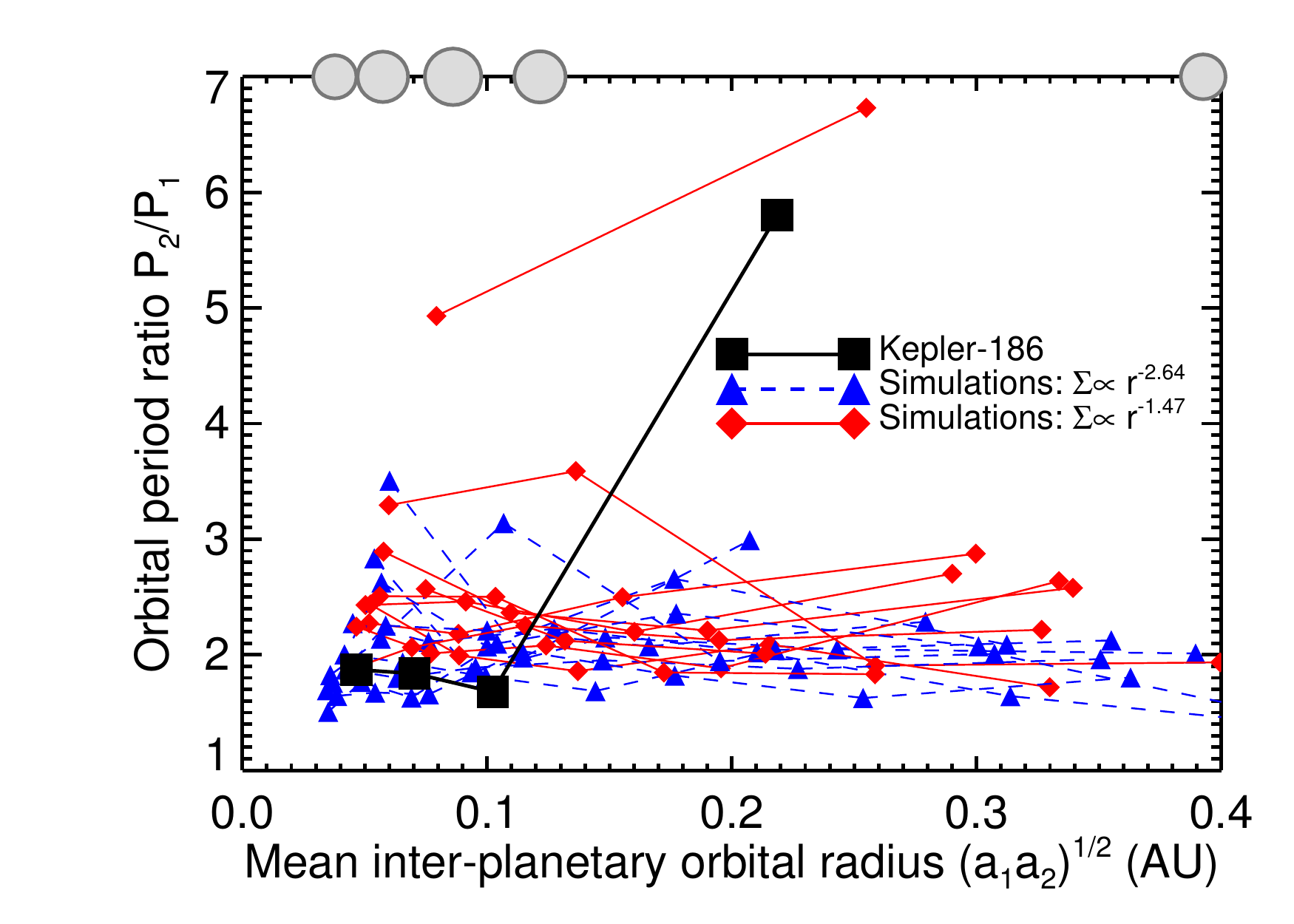}
\caption{Orbital period ratio of pairs of adjacent planets as a function of the two planets' (geometric) mean position. The different sets of simulations are shown with the gray symbols and the real system with the large black squares. Each system is connected. The real system is indicated at the top of the plot. }
\label{fig:prat}
\end{figure}

Figure~\ref{fig:prat} shows the spacing of adjacent planets in the simulated systems compared with the real one. The bulks of simulated planet pairs have period ratios $P_2/P_1$ between 1.5 and 2.5. The planets that formed in the steep ($r^{-2.64}$) disk are more tightly packed, with a median period ratio of $P_2/P_1 = 2.0$, compared with a median of $P_2/P_1 = 2.24$ for the planets that formed in the shallow ($r^{-1.47}$) disk. We attribute this to the larger amount of mass in the inner parts of the disk. This tends to accelerate accretion at early times while there is strong dynamical friction~\citep[see][]{raymond05b}. Given the strong damping from the planetesimal population, growing planets maintain smaller eccentricities and can therefore settle onto more compact orbits than in a dissipation-free environment. In the shallow disk, accretion is slower close-in and the late phases of accretion have less dynamical friction.

For the four inner planets in the Kepler-186 system, the period ratios of adjacent planets $P_2/P_1$ are confined between 1.6 and 1.9.  Farther out is a wide gap: planets e and f have $P_2/P_1 = 5.82$. These two planets are more widely separated than any planet pair in the simulations (except for one exceptional simulation--simulation 6 in Figure~\ref{fig:sims_all}--which only formed three very widely spaced planets). Apart from that case, the most widely separated simulated planet pairs had $P_2/P_1 \approx 3$. One or even two additional planets could comfortably fit between the orbits of planets e and f. All of the simulations contained such planet(s).

Could an additional planet exist between planets e and f but not transit? For that to be the case, that planet would need to have an inclination of at least one to two degrees with respect to the common plane of the other planets (see Section \ref{effectextraplanet}). A collision or scattering event after the dissipation of the gaseous disk could produce such an inclination. It would then be a simple coincidence that planet f's orbit is aligned with the inner ones whereas this extra planet's is not. Or, if this extra planet is somewhat lower-mass than the other planets, then its secular oscillations in inclination could simply reach a higher amplitude than the other planets, decreasing the probability of observing it in common transit with the other planets. 

On the other hand, what conditions of formation would be required for there not to be another planet in between planets e and f? As shown in Figure~\ref{fig:prat}, in situ accretion does not produce large gaps between planets. An alternative is that the planets formed farther out in the disk and migrated inward~\citep{Terquem.etal:2007,cossou13b}. Structure within the disk--such as an opacity transition--can provide a mechanism to stop inward migration at different, or at least time-dependent, orbital radii~\citep[e.g.,][]{masset06,bitsch13,pierens13}. This could, in principle, produce a wide gap between planets. However, any phase of late accretion after migration would likely smooth over such a gap. Indeed, in any planetary system it is difficult to account for large gaps between planets~\citep[e.g.,][]{raymond09c}. 

To conclude this section, we emphasize that local accretion of the Kepler-186 system requires the existence of an additional planet between planets e and f. This applies to both in situ accretion or a late phase of destabilization following inward migration. If there is just one, then an additional planet would likely be located at $\sim$0.2 AU (the geometric mean between planets e and f). 

\subsection{Water Delivery}

Previous work has argued that terrestrial planets orbiting low-mass stars should be relatively dry for two reasons. First, their very short accretion timescales--characterized by high-speed impacts--produce vast amounts of heat that could drive off water~\citep{Lissauer:2007,Raymond.etal:2007b}. Second, given that low-mass stars tend to have lower-mass protoplanetary disks~\citep{Andrews.etal:2007b,Williams.etal:2011}, the lower-mass bodies that grow in these disks do not provide strong enough gravitational ``kicks'' to generate the strong radial mixing needed for water delivery~\citep{Raymond.etal:2007b}.  

The composition of the Kepler-186 planets cannot be strongly constrained from our accretion simulations because the likely source of water is exterior to the simulation domain. If we assume that a division between inner dry material and outer wet material is located at 2.7 AU for a Sun-like star and that this division scales with the stellar flux, then this limit should be located at about 0.5 AU in this system.  In reality, the situation is not that simple.  The inner parts of disks are heated by a combination of stellar irradiation and viscous heating~\citep[e.g.,][]{bitsch13}.  The position of the snow line is a complex function of the stellar and disk properties, both of which evolve in time~\citep[e.g.,][]{kennedy08}.

Our accretion simulations did not consider the outer parts of the Kepler-186 planetary system, which could in principle contain giant planets.  However, gas giants do not play a (positive) role in water delivery~\citep[][]{raymond06a}.  In the classical model of in situ terrestrial planet formation in the solar system, water is delivered to Earth from primitive C-type asteroidal material~\citep{morby00,raymond04,raymond07,raymond14pp6}. Jupiter and Saturn are a hindrance to this process as they eject far more water-rich material than they help to gravitationally diffuse inward toward the terrestrial planets.  It is the disk of solids itself--via gravitational self-stirring--which produces the radial mixing responsible for water delivery.  In the Grand Tack model of terrestrial planet formation, water is delivered to the terrestrial planets by C-type material scattered inward by Jupiter during its outward migration~\citep{walsh11,walsh12,obrien14,jacobson14}.  In both the classical and Grand Tack scenario, Earth would be far more water-rich if it formed in a system with no gas giant~\citep{Raymond.etal:2007b,quintanalissauer2014}.  In the Kepler-186 system, there is no sign of a more distant companion that could hinder water delivery.  We therefore do not think that the outer parts of the system have an important consequence for water delivery in the system.

Are the Kepler-186 planets likely to be wet?  And if so, how wet?  If the Kepler-186 planets formed in-situ (although we consider this to be unlikely; see Section 3.1 above), then water could have been delivered by the disk's gravitational stirring~\citep{raymond08}.  The efficiency of self-stirring depends on the disk mass~\citep{Raymond.etal:2007b}.  Given that the Kepler-186 planets are all as massive or more massive than the solar system's terrestrial planets, the Kepler-186 disk would have been as massive or more massive than the inner parts of the solar Nebula (at least locally).  Thus, gravitational self-stirring--and therefore water delivery--should be as efficient or more efficient in Kepler-186, as in the classical model of terrestrial planet formation~\citep[see][]{raymond14pp6}. Indeed, in our simulations, the feeding zones of the planets generally extend to close to the outer edge of our initial conditions. If the initial conditions were extended to larger orbital radii, then the planets' feeding zones would be wider still. It is likely that each planet's constituent building blocks would thus include water-rich material.ÊBy extrapolating from previous simulations with wider initial conditions and no giant planets \citep{raymond07,raymond08,quintanalissauer2014,RoncoDeElia2014}, we expect that planets near the HZ should accrete a few to 10 percent of their total mass as water-rich bodies. Of course, the impact speeds remain very high in that region and the accretion timescales short~\citep{Lissauer:2007,Raymond.etal:2007b}, so it is unclear how much water would be retained.

On the other hand, if the Kepler-186 planets formed by inward migration, then they should be volatile-rich~\citep{raymond08}. The planets' constituent could have formed with large water content. Their significant masses may have protected them from extensive water loss during giant impacts as well. A measure of the planets' masses and bulk densities to within a few percent is needed to extract information about the bulk water content~\citep{selsis07b}.

\section{Dynamical stability}
\label{stability}

A system of two planets in orbit around a star is dynamically stable if their orbits are separated by at least 3.5 mutual Hill radii $R_{H,m}$~\citep{Marchal.etal:1982,Gladman1993}.  The mutual Hill radius is defined as $R_{H,m} = 1/2 \left(a_1+a_2\right)\left[(m_1+m_2)/3M_\star\right]^{1/3}$, where subscripts 1 and 2 refer to the two planets, $a$ is the orbital semimajor axis, $m$ is the planet mass, and $M_\star$ is the stellar mass.  A system of many planets must be more widely separated than a critical limit of 5--10 mutual Hill radii to ensure long-term stability~\citep{Chambers.etal:1996,Marzari2002}.  

The masses of the \starname planets are of course unknown.  Table~\ref{tab:plcomp} lists the planets' masses for the widest plausible range in compositions, calculated with the mass--radius relations of \cite{Fortney.etal:2007}.  We assumed that the planets are solid and do not contain enough H/He gas to alter their radii \citep{Weiss2013,Weiss2014}.     

The dynamical inter-planet spacing depends on the planets' true masses.  Planets d and e are dynamically closest together and planets e and f are the most widely spaced.  Lower-density, lower-mass planets are more widely spaced in dynamical terms (mutual Hill radii).  For pure ice, planets d and e are separated by $15 R_{H,m}$, but this value decreases with increasing planet mass to $9 R_{H,m}$ for Earth-like compositions and just $6.5 R_{H,m}$ for pure iron planets.   The gap between planets e and f is wide enough to fit another planet.  For ice-rock-Earth-iron planets, the gap is 55-37-34-25 mutual Hill radii wide.  It is therefore not surprising that our accretion simulations formed extra planets in this region.  

The \starname system is dynamically stable.  We ran a suite of $N$-body simulations of the five-planet system for the full range of planetary compositions.  Given the weak constraints on the planets' eccentricities and longitudes of pericenter, we sampled a range of orbital phases and included initial eccentricities up to 0.05.  In all cases the systems were stable for the 0.1~Myr duration.  We ran 10 longer-term simulations (without tides or general relativity) with pure iron planets, all of which were stable for 100~Myr.

\section{Tidal orbital evolution}
\label{tidal}

Given the proximity of the system to its star, tidal interactions are important in shaping the long-term dynamical evolution of the system. Tides affect a close-in planet's orbit in several ways. On short timescales, they drive the system to an equilibrium rotation state, typically either a spin-orbit resonance or a ``pseudo-synchronous'' state whereby the planet corotates with the star at its closest approach~\citep{Hut:1981,Ferrazmello.etal:2008,Makarov.etal:2013}. Dissipation within the planet decreases the planet's eccentricity and obliquity. Changes in orbital distance, driven by dissipation in the planet or star, occur on longer timescales. 

In multiple-planet systems with close-in planets, the orbital evolution is a combination of eccentricity pumping from planet--planet gravitational forcing and damping from tidal interactions~\citep{Mardling:2007,Bolmont.etal:2013}. The strength of eccentricity pumping is determined by the planets' masses and orbits and the degree of tidal dissipation by the (unconstrained) dissipation rates, especially the planet with the strongest dissipation (usually but not always the closest-in one). 

We simulated the long-term dynamical and spin evolution of the \starname system. Our simulations included the tidal dissipation model of \cite{Hut:1981} and \cite{leconte2010} a post-Newtonian precession term~\citep{kidder1995}. Both were applied to all five planets.  The code is a fully three-dimensional (3-D) version of the one used in \citet{Bolmont.etal:2013}, it now computes the tidal evolution for planets on inclined orbits and with a non-zero obliquity. Most of the work done so far considered coplanar systems and did not compute the evolution of the obliquity of the planets \citep[e.g.,][]{DobbsDixon2004,Mardling:2007,Batygin2009}. We also added the effect of the rotation-induced flattening of the star and of the planets \citep[such as in][]{Murray.etal:1999,CorreiaRodriguez2013}. 

\subsection{Exploring the Planets' Mass Range}

In order to have a vision as broad as possible on the dynamical evolution of the \starname system, and since the masses of the planets are not constrained, we performed simulations assuming various compositions for planets. 

We first tested the extremes: 100\% ice planet and 100\% iron planets. Both cases are very unlikely if not impossible but they allow us to investigate the evolution of the system for very low mass planets and very high mass planets. We also tested some intermediate compositions: 50\% ice- 50\% rock and Earth-like composition. The corresponding masses are stated in Table \ref{tab:plcomp}.

We assumed that the 100\% ice planets have a dissipation $k_2 \Delta t$ higher than that of the Earth\footnote{$k_2$ is the Love number of degree 2 and $\Delta t$ is the time lag~\citep[see][]{Hut:1981}. $k_2 \Delta t$ is assumed constant. For Earth, $k_{2,\oplus} \Delta t_\oplus = 213$~s \citep{DeSurgyLaskar1997}.} \citep[e.g.,][]{McCarthyCastillo2013}. We also assume that the 100\% iron planets have a dissipation $k_2 \Delta t$ higher than that of the Earth \citep[e.g.,][]{KootDumberry2011,JacksonFitzGerald2000}. For these two compositions, we tested 1 and 10$\times~k_{2,\oplus} \Delta t_\oplus$. For intermediate compositions, we tested 0.1 and 1$\times~k_{2,\oplus} \Delta t_\oplus$. Earth's dissipation is quite high due to the friction of shallow water reservoirs on the crust \citep[e.g.,][]{Lambeck1977}. For a different planet--with no water or different topology--this efficient dissipation mechanism might be absent; therefore, in order to bracket what might be an appropriate value, we consider that intermediate composition planets have a dissipation rate in a range from 0.1 to 1$\times~k_{2,\oplus} \Delta t_\oplus$. The planets were given randomly chosen orbital angles, initial eccentricities of less than 0.06, initial inclinations of less than 0.4 degree\footnote{We choose small initial inclinations so as to ensure that \planetf transits.} and obliquities less than $30$ degree.

\begin{figure}
\includegraphics[width=0.45\textwidth]{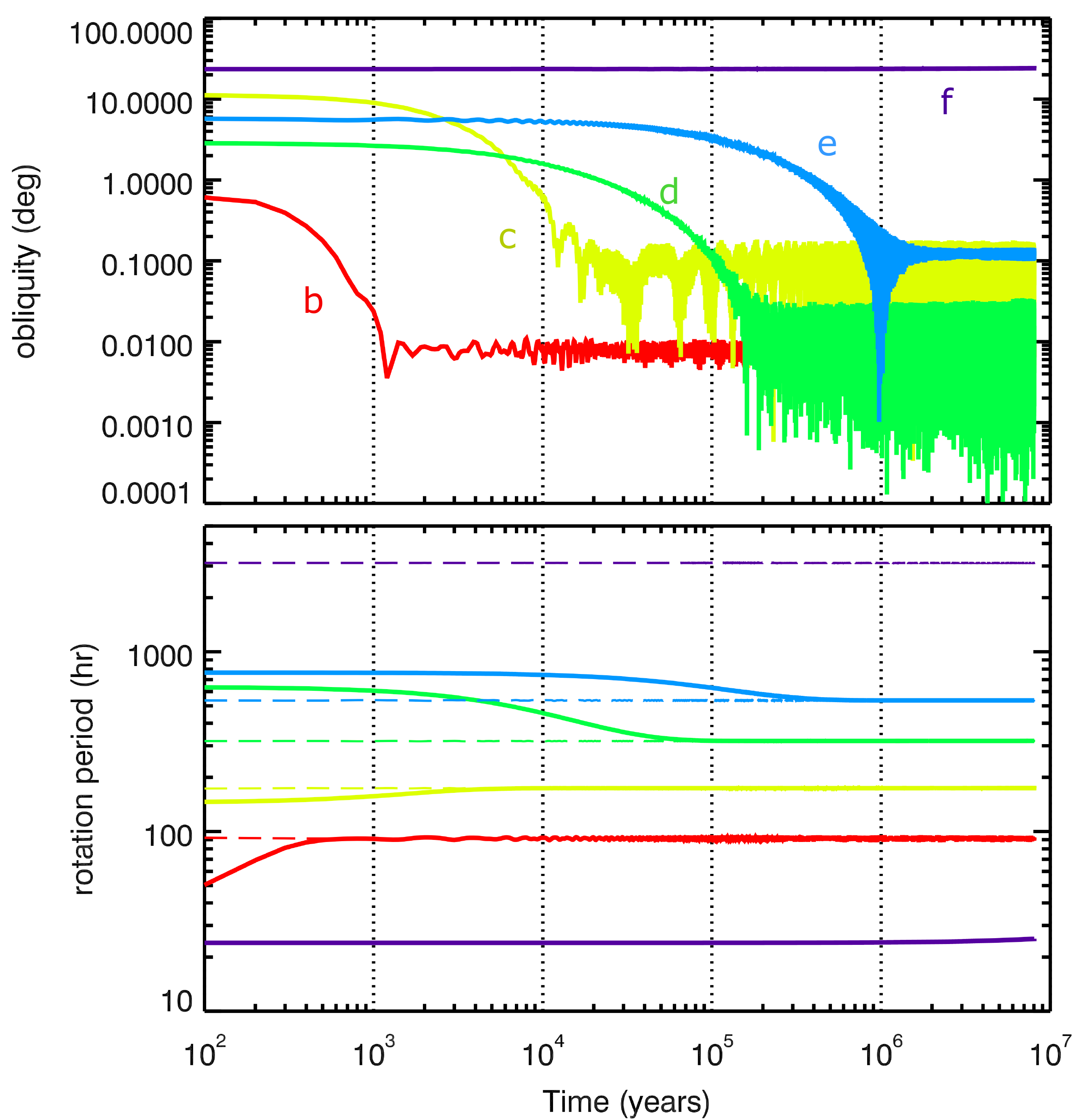}
\caption{Evolution of the five planets' obliquities (top) and rotation periods (bottom) for set $\mathcal{A}$. In the bottom plot, the solid lines correspond to the planets' actual spin periods and the dashed lines to the pseudo-synchronous values. Given their small orbital eccentricities, the pseudo-synchronous rotation states are very close to 1:1 spin--orbit synchronous rotation. The timescale for tides is strongly dependent on the orbital radius. The ordering of the planets is clearly discernible from the order in which their obliquities decay.}
\label{fig:obl-rot}
\end{figure}

Figure~\ref{fig:obl-rot} shows the evolution of the planetary spins during a 20~Myr simulation, assuming Earth-like compositions for all the planets and $k_2 \Delta t$ values equal to Earth's. The timescale for tidal interactions is short enough for the four inner planets' evolution to be significantly affected by tides on megayear timescales. The obliquities of the four inner planets were reduced to nearly zero within $\sim$1 Myr, regardless of composition and tidal dissipation. Likewise, the rotation rates of the four inner planets converged to their pseudo-synchronous values (dashed lines). Given their small eccentricities, this means that the planets rotate extremely slowly, effectively synchronously. \planetf is evolving towards pseudo-synchronization, and due to the short initial rotation period its obliquity starts increasing. On the long term, the rotation period of \planetf lengthens and its obliquity starts to decrease toward its (near zero) equilibrium value. 

\begin{figure}
\includegraphics[width=0.45\textwidth]{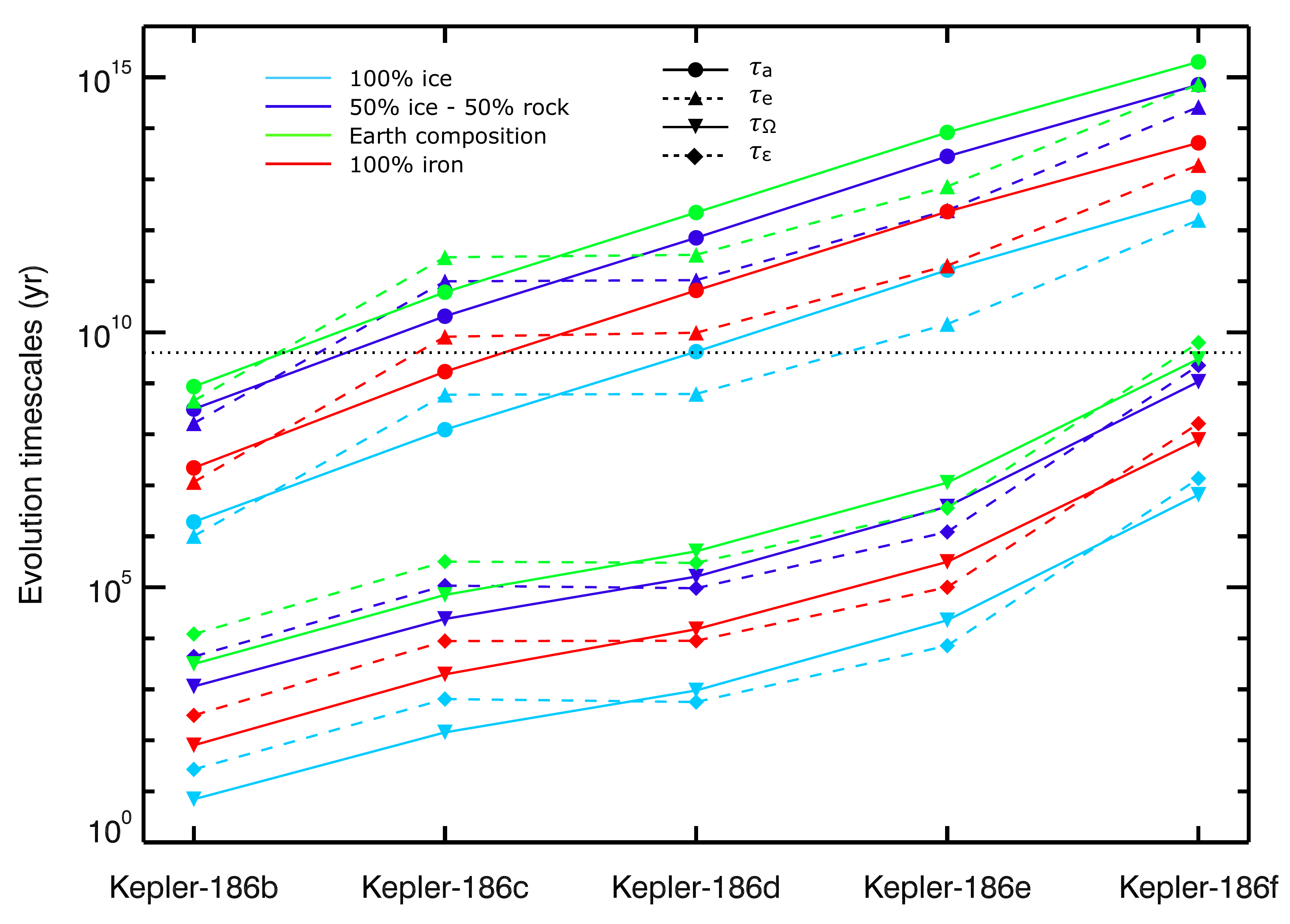}
\caption{Timescales for planetary tide-induced evolution for the Kepler-186 system and for the four compositions considered here. $\tau_a$, $\tau_e$, $\tau_\Omega$, and $\tau_\epsilon$ are, respectively, the timescales of evolution of semi-major axis, eccentricity, rotation rate, and obliquity. The horizontal black dotted line corresponds to the estimated system age of 4~Gyr. The timescales were computed here for set $\mathcal{A}$, however, the timescales of evolution for set $\mathcal{B}$ are of the same order of magnitude.}
\label{timescales}
\end{figure}

Figure \ref{timescales} shows the system's tidal evolution timescales for set $\mathcal{A}$. The timescales are shown for each planet and for the different compositions assuming the lower values of $k_2 \Delta t$: $1,0.1,0.1,1\times k_{2,\oplus} \Delta t_\oplus$ for 100\% ice, 50\%--50\% rock ice, Earth composition, and 100\% iron planets, respectively. A combination of lower masses and rather high dissipation rates makes the tidal evolution timescales much shorter for pure ice planets than for the other compositions. The evolution timescales computed for set $\mathcal{A}$ are of the same order of magnitude as those computed with set $\mathcal{B}$.

Constraining the age of the system could possibly allow us to constrain the compositions of the planets. Indeed, the timescale for the semi-major axis evolution of \planetb is quite short--assuming a composition of 100\% ice and 100\% iron--meaning that such a planet would be falling on its host star in timescales that are probably shorter than the system lifetime. It is therefore more probable that the composition of \planetb, and also of \planetc, is rocky.

The evolution timescales of the semi-major axis and the eccentricity of \planete and \planetf are higher than the age of the universe. For rocky compositions, the evolution timescales of the semi-major axis and eccentricity of \planetc and \planetd are also longer than the age of the universe. We then expect the four outer planets to have been formed about where they are now located. However, \planetb is likely to have been formed a bit further away.

\subsection{Influence of Eccentricity}

The eccentricities of the planets are poorly constrained. In order to obtain the values used in the previous section, we used the median values of \citet{Quintana2014}. We also simulated the evolution of the system for higher eccentricities, within the range allowed by the observations. The maximum eccentricities are, from b to f, $\sim0.3$, $0.3$, $0.3$, $0.3$, and $0.4$.

For each planetary composition, from 100\% ice to 100\% iron, assuming the maximum eccentricities for each planet leads to a destabilization of the system in less than 1000~yr.

Given that the timescale of the evolution of the eccentricity of \planetb is lower than 1~Gyr, we can assume that its eccentricity is low. The eccentricity of \planetb should have the equilibrium value obtained by competition between tidal damping and the excitation due to planet--planet interactions. We assumed it to be 0.05. In order to have non-crossing orbits, this means that the eccentricity of \planetc has to be lower than 0.3, the eccentricity of \planetd has to be lower than 0.13, the eccentricity of \planete lower than 0.2, and the eccentricity of \planetf lower than 0.6. For \planetd and e, this slightly reduces the eccentricity range consistent with the observations. However, it does not constrain the eccentricity of \planetc and \planetf.
 
We explored only a part of the huge parameter space for the eccentricities in order to have a general idea of the stability of the system for different initial eccentricities. Most of the time, the destabilization concerns \planetb and \planetc. The dynamics of the system is influenced by the massive planets \planetc and \planetd. In order to increase the stability of the system, we had to consider eccentricities for these two planets to be relatively low.

We found some configurations stable for at least 1~Myr with an Earth-like composition planets and Earth-like dissipation factor, with initial eccentricities for \planetb to \planetf of 0.05, 0.1, 0.1, 0.1 and 0.2. However, this configuration leads to an excitation of the inclination of the four inner planets after $\sim2\times10^5$~yr of evolution to a level superior to their limit inclinations ($\arctan(R_\star/a)$, where $a$ is the planet semi-major axis). The four inner planets spend, respectively, 52\%, 53\%, 50\%, and 41\% of the simulation time out of transit configuration. It is therefore unlikely that the planets have such high eccentricities.

For planets which are 100\% ice, the system can be stable over 1~Myr for initial eccentricities of \planetb to f of 0.15, 0.14, 0.1, 0.1, and 0.4. The inclinations in this configuration are consistent with transit. The eccentricity of \planetf would be unconstrained, as the masses of the planets being low, it is dynamically independent. However, for denser compositions, \planetf can have some influence over time on the four inner planets and lead to a slow increase of their eccentricities which leads to a destabilization.

Assuming that the planets are rocky, which is maybe the most probable configuration, this study shows that the eccentricities of the planets cannot be too high (typically, they have to be inferior to $\sim0.08$ for the four inner planets). The eccentricity of \planetf can be as high as 0.2.

\subsection{The Effect of an Extra Planet on the System's Dynamics}\label{effectextraplanet}

An extra planet could exist in the \starname system between \planete and \planetf (see Section 3.3). We therefore simulated the dynamical evolution of the system, adding an inclined extra planet in order to see its influence on the observable planets. We performed these simulations assuming Earth-like compositions and a $k_2 \Delta t = 0.1\times~k_{2,\oplus} \Delta t_\oplus$ for planets b, c, d, e and f. 

We considered an extra planet with mass between $0.1M_\oplus$ and $1 M_{Jup}$. The rocky planets (from 0.1 to 10~$M_\oplus$) have a $k_2 \Delta t$ of $0.1\times~k_{2,\oplus} \Delta t_\oplus$. The Neptune mass planet has a $k_2 \Delta t$ of 0.038~s and the Jupiter mass planets has a $k_2 \Delta t$ of $7\times 10^{-5}$~s. The extra planet has a semi-major axis of $0.233$~AU, an initial eccentricity of 0.01 and an initial inclination of 2$\deg$ so as not to transit. Its initial obliquity is $17^\circ$ and its initial rotation period is $24$~hr. 

The accretion simulations in Section \ref{formation} tell us that the mass range should probably be narrower but we chose here not to constrain the parameter space. 

Table \ref{tab:proba} shows a measure of the observability of the planets in the simulations: the transit probability is the fraction of the simulation during which each planet has an inclination lower than $\arctan(R_\star/a)$, where $a$ is the planets' semi-major axis. The values of the inclination above which the transit is geometrically impossible for each planet (b, c, d, e, extra planet, and f): 3.5, 2.3, 1.5, 1.1, 0.60 and 0.34 degree for set $\mathcal{A}$ and 3.3, 2.2, 1.5, 1.0, 0.58, and 0.32 degree for set $\mathcal{B}$.

\begin{table*}
\begin{center}
\caption{Probability of Transit} 
\vspace{0.1cm}
\begin{tabular}{c|cccccc}        
\hline         
Mass of Extra & \multicolumn{6}{c}{Prob. of Transit for Sets $\mathcal{A}$ and $\mathcal{B}$ (Calculated for a 20~Myr Simulation)} \\  
Planet & \planetb & \planetc & \planetd & \planete & Extra Planet & \planetf \\
 & Set $\mathcal{A}$--Set $\mathcal{B}$ & Set $\mathcal{A}$--Set $\mathcal{B}$ & Set $\mathcal{A}$--Set $\mathcal{B}$ & Set $\mathcal{A}$--Set $\mathcal{B}$ & Set $\mathcal{A}$--Set $\mathcal{B}$ & Set $\mathcal{A}$--Set $\mathcal{B}$ \\
\hline
0.1~$M_\oplus$  & 100\%--100\% & 100\%--100\% & 100\%--100\% & 100\%--100\% & 0\%--0\% & 82\%--89\% \\
1~$M_\oplus$     & 100\%--100\% & 100\%--100\% & 100\%--100\% & 100\%--100\% & 0\%--0\% & 37\%--33\% \\
10~$M_\oplus$   & 100\%--95\% & 82\%--68\% & 51\%--46\% & 34\%--30\% & 18\%--13\% & 4\%--4\% \\
1 $M_\mathrm{Neptune}$   & 85\%--74\% & 56\%--49\% & 40\%--38\% & 24\%--19\% & 0\%--0\% & 3\%--3\% \\
1 $M_\mathrm{Jupiter}$& 55\%--46\% & 42\%--44\% & 24\%--27\% & 17\%--17\% & 0\%--0\% & 5\%--5\% \\
\hline
\end{tabular} 
\label{tab:proba} 
\end{center}
\end{table*}

Due to the initial small eccentricity of the extra planet, the eccentricities of planets b, c, d, e, and f are not excited to levels incompatible with the observations. In particular, the eccentricity of \planetf remains always below 0.03. However, the inclination is excited by the presence of the extra planet and the more massive the extra planet the higher the other planets' inclinations.

\begin{figure}
\includegraphics[width=0.45\textwidth]{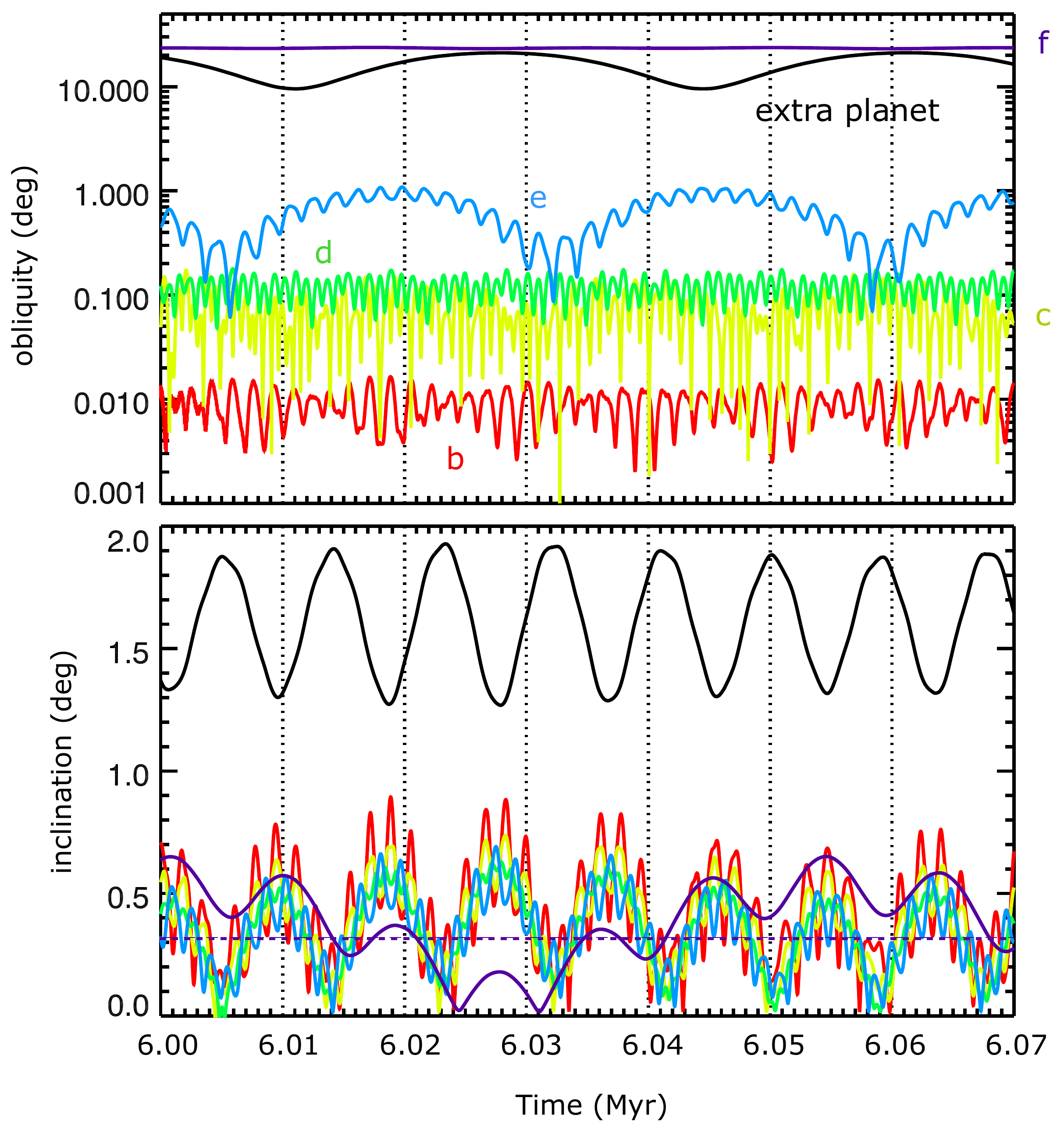}
\caption{Snapshot of the evolution of the obliquities and inclinations of the six planets over 70,000 yr (set $\mathcal{B}$). The colored lines correspond to the five confirmed planets (from red to purple: from b to f), and the black line corresponds to the hypothetical extra planet (1~$M_\oplus$). In the bottom plot, the purple dashed line corresponds to the limit inclination over which \planetf does not transit.}
\label{14363_o}
\end{figure}

Adding a massive non-transiting planet increases the mutual inclinations of the other planets, and thus decreases the probability of a transit of \planetf considerably: it decreases from 88\% when the extra planet is 0.1~$M_\oplus$ to 33\% when the planet is 1~$M_\oplus$. It is therefore unlikely that the system hosts a planet more massive than 1~$M_\oplus$ between \planete and \planetf. Figure \ref{14363_o} shows the evolution of the system with an extra 1~M$_\oplus$ planet after 6~Myr of integration. The bottom plot shows the limit inclination over which \planetf does not transit (dashed purple line). The inclination of \planetf oscillates and sometimes becomes greater than the limit $\arctan(R_\star/a_{\rm f})$.

Adding a planet in the system allows angular momentum to transfer to the outer planet much more efficiently and this has an influence on its obliquity. Figure \ref{14363_o} shows the evolution of the system with an extra 1~$M_\oplus$ planet. We can see that instead of having a purely tidal evolution, the obliquity of \planetf oscillates between 23$\deg$ and 24$\deg$ with a main frequency of $\sim10^4$~yr. If the extra planet is 10~$M_\oplus$, then the obliquity of \planetf oscillates between 18$\deg$ and 24$\deg$. Oscillations of the planet's obliquity would have an influence on the planet's climate~\citep[e.g.,][]{Armstrong2014}. 

An extra 1~$M_\oplus$ planet also affects the equilibrium obliquities of the four inner planets. The influence is greater for \planete and \planetd, but they all stay below 1$\deg$. 

When the mass of the extra planet is larger, the obliquities of the planets are higher and can reach values of a few degrees. Tides are less efficient to counteract the excitation due to the high mass extra planet. 

\subsection{Evolution of \planetf}

With no extra planet in the system, \planetf is dynamically isolated from the four inner planets. Its orbit and spin evolve due to the tides it raises on the star and those raised in it by the star rather than gravitational interactions with other planets. 

\begin{figure}
\includegraphics[width=0.45\textwidth]{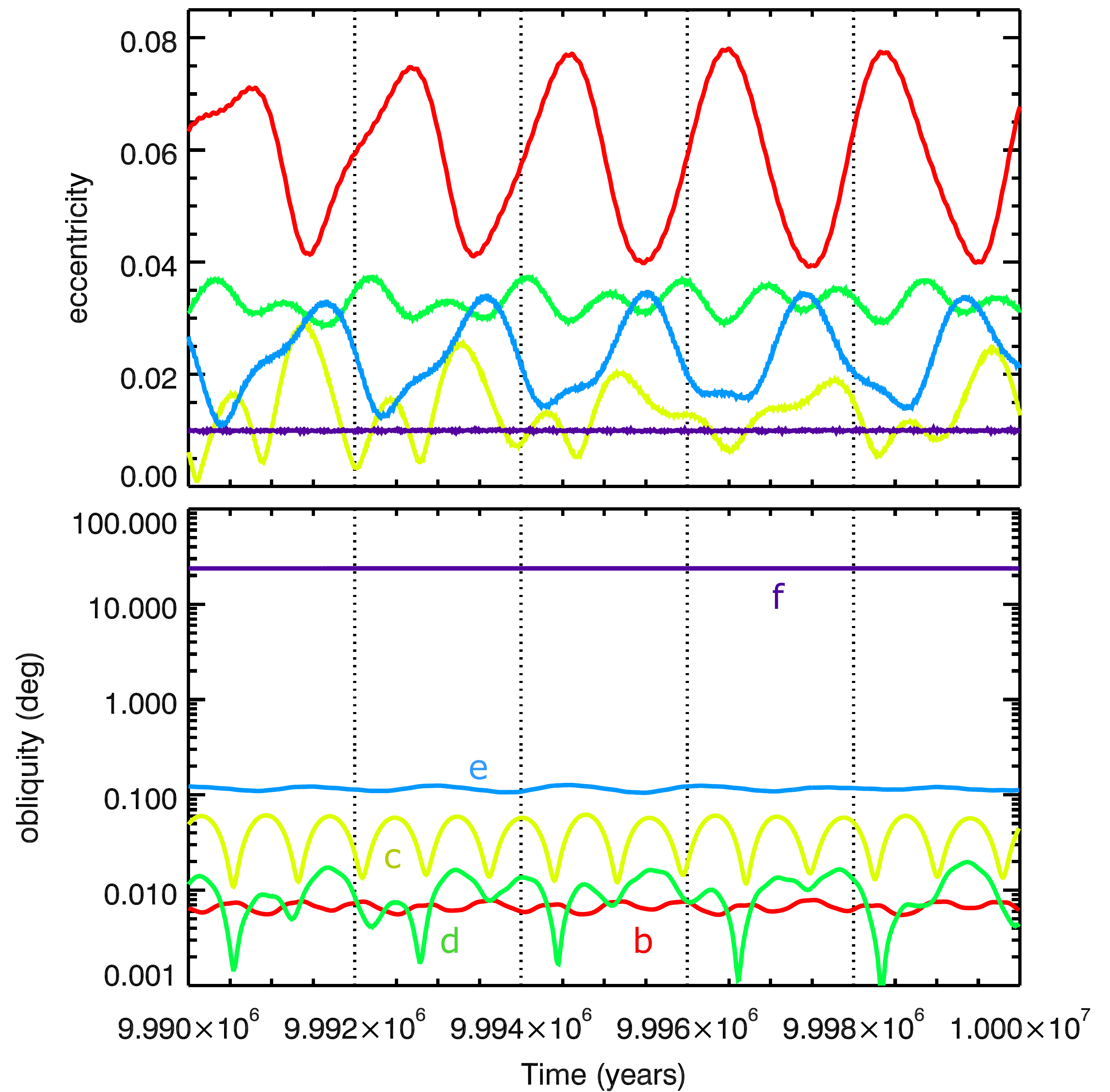}
\caption{Short-term (10,000 yr) evolution of the eccentricities (top) and obliquities (bottom) of the five planets in the \starname system (set $\mathcal{B}$).}
\label{fig:ecc-obl}
\end{figure}

Figure~\ref{fig:ecc-obl} shows the evolution of the eccentricities and obliquities of the five planets over the last 5000 yr of the simulation from Figure~\ref{fig:obl-rot}. The eccentricity and obliquity of \planetf do not undergo noticeable oscillations, whereas the four inner planets' do. Their eccentricities oscillate as a combination of frequencies which correspond to the secular modes of the system~\citep[see, for example,][]{Murray.etal:1999}. The amplitudes of oscillation are a few percent and the characteristic secular timescales are $\sim$1000 yr. The oscillation amplitudes are mass-independent but the frequencies increase linearly with the planet masses. Oscillations in eccentricity can cause a modest change in the insolation received by a planet, as the orbit-averaged insolation scales with eccentricity $e$ as $\left(1-e^2\right)^{-1/2}$. This in turn can, in some instances, trigger changes in the planetary climate on the secular timescale~\citep{Spiegel.etal:2010}. Indeed, large climatic events are thought to correlate with oscillations of Earth's orbital quantities, especially its eccentricity and obliquity~\citep[so-called Milankovitch cycles;][]{berger1988}. 

Figure \ref{timescales} shows that the planetary tide does not cause the eccentricity and semi-major axis of \planetf to evolve on timescales shorter than 10~Gyr and the stellar-tide induced evolution occurs on timescales even longer: $>10^{17}$~yr. However, the evolution timescales for the obliquity and rotation period are of the order of magnitude of 1~Gyr for rocky compositions. Therefore, given the age of the system, \planetf could have reached pseudo-synchronization and very low obliquity or could still be evolving. 

Figure~\ref{fig:obl-rot} shows the very slow tidal evolution of \planetf. Over 20 Myr \planetf retains its initial obliquity, rotation rate, and eccentricity. However, toward the end of the simulation, \planetf's rotation appears to be slowly decreasing and its obliquity slowly increasing. 

We calculated several possible long-term evolutionary pathways for \planetf's spin state. The initial obliquity was varied from Earth's current obliquity of $23.5\deg$ to an obliquity of 80$\deg$. The initial spin rate was varied over the range $8.7\times 10^{-7}$--$3.5\times10^{-4}$~s$^{-1}$, which correspond to rotation periods from 5~hr to 2000~hr. Here, the planet is assumed to have the same dissipation factor $k_2 \Delta t$ as Earth. Unlike the simulations from Figures~\ref{fig:obl-rot} and~\ref{fig:ecc-obl}, these calculations were performed with only \planete and \planetf in the system. They were nonetheless a fully 3D implementation of the {\em constant time lag} equilibrium tidal model~\citep{Hut:1981,leconte2010}. 

\begin{figure}
\includegraphics[width=0.45\textwidth]{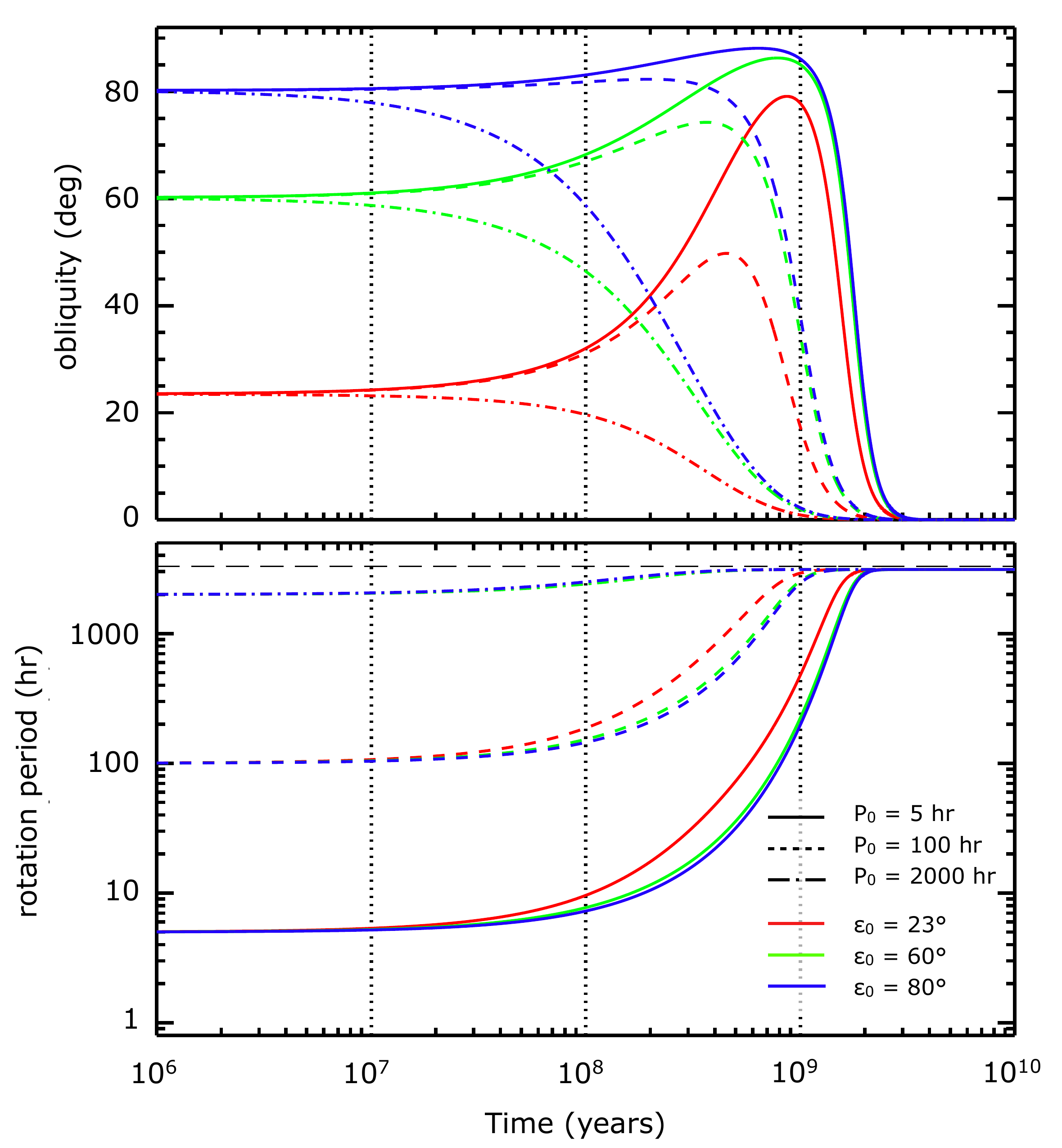}
\caption{Long-term evolution of the obliquity (top) and rotation period (bottom) of \planetf (set $\mathcal{A}$). Each set of linestyle curves represents a different initial spin rate and each set of colored curves represents a different initial obliquity. The thick, black dashed line represents the pseudo-synchronous rotation which, for this zero-eccentricity example, is the 1:1 spin--orbit resonance. The fastest-spinning (red full line) curve is closest to the example shown in Figure~\ref{fig:obl-rot}. This evolution was computed with set A.}
\label{fig:obl_f}
\end{figure}

Figure~\ref{fig:obl_f} shows that planet f's obliquity increases for all but the slowest initial spin rate. The period of obliquity increase lasts for a few hundred megayears. It is followed by a long, slow decay that lasts 2-3 Gyr (for the arbitrarily chosen range of initial spin rates), which is broadly consistent with the results of \cite{heller2011}. The initial obliquity and spin rates are of course unknown, although $N$-body simulations of terrestrial accretion produce planets with fast initial spins and  isotropically distributed obliquities~\citep{kokubo07}. 

Evolution timescales scale inversely with the planet's dissipation rate, assumed here to be roughly Earth-like. Stronger dissipation accelerates the evolution whereas weaker dissipation slows it down. The age of \starname is thought to be higher than a few gigayears, so assuming an Earth-like dissipation would mean that \planetf should be in a pseudo-synchronous rotation state with small obliquity. However, the age of the system is unconstrained. So if \starname is somewhat younger (say 1 Gyr) or if the dissipation within planet f is inefficient,\footnote{It is likely that the dissipation of \planetf is actually lower than that of the Earth, so the curves of Figure \ref{fig:obl_f} could be shifted to the right, meaning that the state of pseudo-synchronization and low obliquity occurs later.} then \planetf should not have reached a pseudo-synchronous state. In that case, although \planetf's spin rate would probably have slowed to within a factor of a few of the pseudo-synchronous rate, its obliquity would be unconstrained and could assume very high values ($\sim$85$\deg$ for the full blue line of Figure \ref{fig:obl_f}).

The heat flux generated by the deformation of a planet can influence its atmospheric properties and climate \citep[e.g.,][]{barnes2009,Barnes2013}, so we also investigated the effect of tidal heating on the thermal history of \planetf. Our simulations show that the tidal heat flux could be sustained at values of more than 0.1~W/m$^2$ for at least the first 10~Myr of evolution (assuming Earth composition, $k_{2,\oplus} \Delta t_\oplus$), more than 1~W/m$^2$ for the first 10$^5$ yr of evolution (assuming pure ice composition, 10$\times k_{2,\oplus} \Delta t_\oplus$), and as high as 4.7~W/m$^2$ for an eccentric configuration of the system (assuming pure ice composition, 10$\times k_{2,\oplus} \Delta t_\oplus$).
However, atmospheric modeling results from Section \ref{habi} suggest that even for dense atmospheres, the radiative flux at the surface exceeds the tidally induced fluxes by at least a factor of 10.

\section{Habitability of Kepler-186f}
\label{habi}

Applying approximate equations from \citet{kopparapu2013,kopparapu2014} for the Kepler-186 system results in an approximate inner boundary of the HZ (runaway greenhouse limit) of 0.20--0.23\,AU, depending on the adopted stellar parameters and the choice of parameters (either \citealp{kopparapu2013} or \citealp{kopparapu2014}). The corresponding outer boundary of the HZ (maximum greenhouse limit) ranges from 0.40--0.43\,AU. A more optimistic empirical estimate of the width of the HZ \citep{Selsis2007} yields a range from 0.15--0.42\,AU, taking into account the possible effect of clouds near the HZ boundaries. This suggests that within the uncertainties of its orbital distance, Kepler-186f is indeed in the HZ.

In terms of insolation, Kepler-186f receives $S_{K186}$=0.32$^{+0.05}_{-0.03}$ times the insolation as the present Earth \citep{Quintana2014}. Note that applying consistent sets of parameters (stellar luminosity, orbital distance, etc.) yields slightly lower insolation values for Kepler-186f (see Table \ref{tab:planet}) than stated in \citet{Quintana2014}. However, within 1$\sigma$ uncertainty, this is the same insolation ($S_{G581}$=0.29) as that received by the super-Earth candidate GJ 581d, which climate models have shown to be capable of having liquid water on its surface \citep[e.g.,][]{wordsworth2010,vparis2010gliese,hu2011,kaltenegger2011,wordsworth2011}, given a large enough CO$_2$ greenhouse effect ($p_{CO_2}\gtrsim$1--2\,bar). 

For a preliminary assessment of the habitability of Kepler-186f, we used the one-dimensional, cloud-free radiative-convective atmosphere model from \citet{vparis2010gliese}. Model atmospheres were assumed to be composed of CO$_2$, N$_2$, and H$_2$O only. N$_2$ and CO$_2$ are assumed to be well-mixed throughout the atmosphere. H$_2$O concentrations are calculated following the ambient temperature (hence, vapor pressure) and a prescribed relative humidity profile \citep{manabewetherald1967}. We performed a series of calculations, varying planetary gravity and insolation, as well as N$_2$ and CO$_2$ partial pressures.  Input parameters are listed in Table \ref{input_data}. The assumed range of CO$_2$ and N$_2$ partial pressures is plausible if Kepler-186f is a rocky planet. The volatile budget of Earth is thought to consist of several bars of N$_2$ and tens to hundreds of bars of CO$_2$ \citep[e.g.,][]{turekian1975,kasting1988,mckay1989mars,lundin2004,goldblatt2009faintyoungsun}. The needed stellar input spectrum for the model simulations was calculated from a synthetic spectrum using stellar models from \citet{hauschildt1999}, stellar parameters from Table \ref{tab:star}, and a metallicity of --0.28 \citep{Quintana2014}.

\begin{figure*}[h]
  \centering
      \includegraphics[width=520pt]{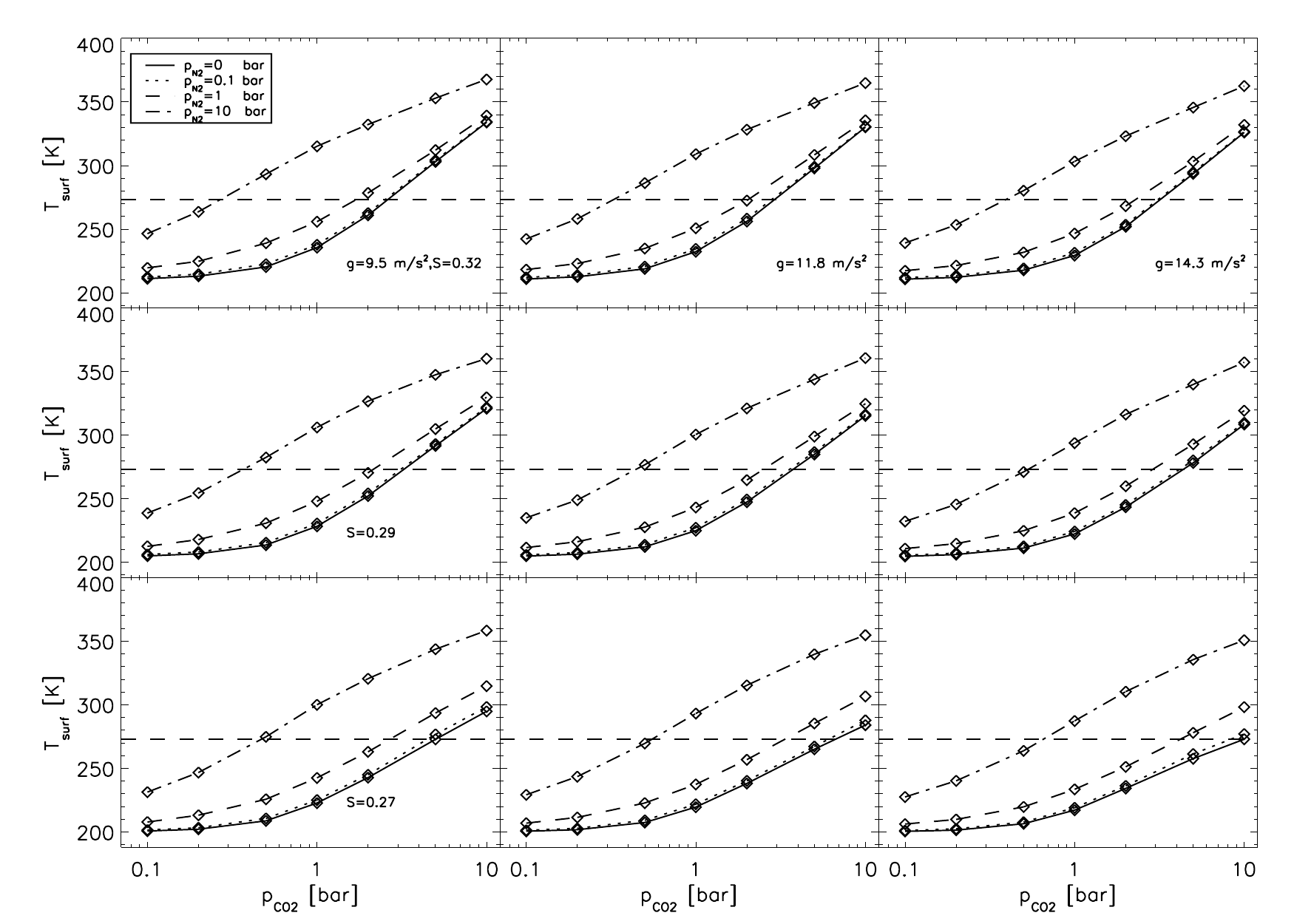}\\
  \caption{Surface temperature as a function of CO$_2$ partial pressure for different N$_2$ partial pressures. Water triple-point temperature of 273\,K indicated by horizontal dashed line.  Top to bottom rows: decreasing insolation (from top to bottom: \citealp{Quintana2014}, set $\mathcal{A}$, set $\mathcal{B}$ from Table \ref{tab:star}). Left to right columns: increasing gravity.}\label{contourplots}
\end{figure*}

Figure \ref{contourplots} shows the calculated surface temperatures as a function of CO$_2$ partial pressure for different N$_2$ partial pressures.  These results clearly suggest that \starname f is a potentially habitable planet if it is a rocky planet and Earth-like in bulk composition. To reach mean surface temperatures above freezing, modest amounts of CO$_2$ are needed for most of the cases. For a large atmospheric reservoir of N$_2$, surface temperatures rise above 273\,K already at about 200--500\,mbar of CO$_2$, again for almost every scenario. 

Furthermore, the results shown in Figure \ref{contourplots} imply a strong influence of N$_2$ on the calculated surface temperatures, consistent with results in previous studies \citep[e.g.,][]{Li2009,goldblatt2009faintyoungsun,vparis2010gliese,wordsworth2010,vparis2013marsn2,kopparapu2014}. High N$_2$ partial pressures could reduce the amount of CO$_2$ needed to maintain a surface temperature above freezing by almost an order of magnitude.

As can be inferred from Figure \ref{contourplots}, a change in gravity due to uncertainties in stellar radius, transit depth, and planetary mass (Tables \ref{tab:star}-\ref{tab:plcomp}) has a modest influence on the calculated surface temperatures. With increasing gravity, surface temperatures decrease by 5-10\,K. This is mainly due to a decrease in column density (at fixed pressure), consistent with previous studies \citep[e.g.,][]{wordsworth2010,wordsworth2011,rauer2011}.  Generally, assuming that \starname f is indeed a rocky planet, mass and radius estimates, and hence planetary gravity, are not found to be critical for habitability. 

From Table \ref{tab:star} \citep[or, e.g., Table 1 in][]{borucki2011}, it is apparent that the stellar mass and radius are not constrained at very high precision (partly due to the faintness of \starname). Since stellar mass directly impacts planetary orbital distance (via Keplers Third Law) and stellar radius is related to the luminosity, insolation for \starname f varies by as much as 20\% for sets $\mathcal{A}$ and $\mathcal{B}$ or orbital distances as stated in \citet{Quintana2014}. Naturally, this has a certain impact on the calculated surface temperatures, as shown in Figure \ref{contourplots}. Upon increasing stellar insolation, surface temperatures increase by  10-60\,K, depending on CO$_2$ and N$_2$ partial pressure. This is a much larger effect than for gravity. Taking for example Set $\mathcal{B}$ at high gravity (lower right in Fig. \ref{contourplots}), a minimum of 1\,bar N$_2$ is required to reach habitable surface temperatures even when assuming 10\,bar of CO$_2$.  Hence, our results emphasize the need for accurately determined stellar parameters for habitability studies.

Calculated surface temperatures in Figure \ref{contourplots} rise up to 350--370\,K for high-pressure atmospheres. Such high temperatures are not conducive for higher lifeforms on Earth, although extremophiles are known that can thrive under these conditions \citep{rothschild2001}. Another potential challenge for lifeforms might be the increased pH value of rain due to high amounts of atmospheric CO$_2$ \citep[e.g.,][]{ohmoto2004}, which again is tolerated in principle by some extremophile species on Earth \citep{rothschild2001}. Therefore, it is at least conceivable from an Earth-centric view that a microbial biosphere could exist under the atmospheric and surface conditions calculated for \starname f. 

\begin{figure}[h]
  \centering
 \includegraphics[width=250pt]{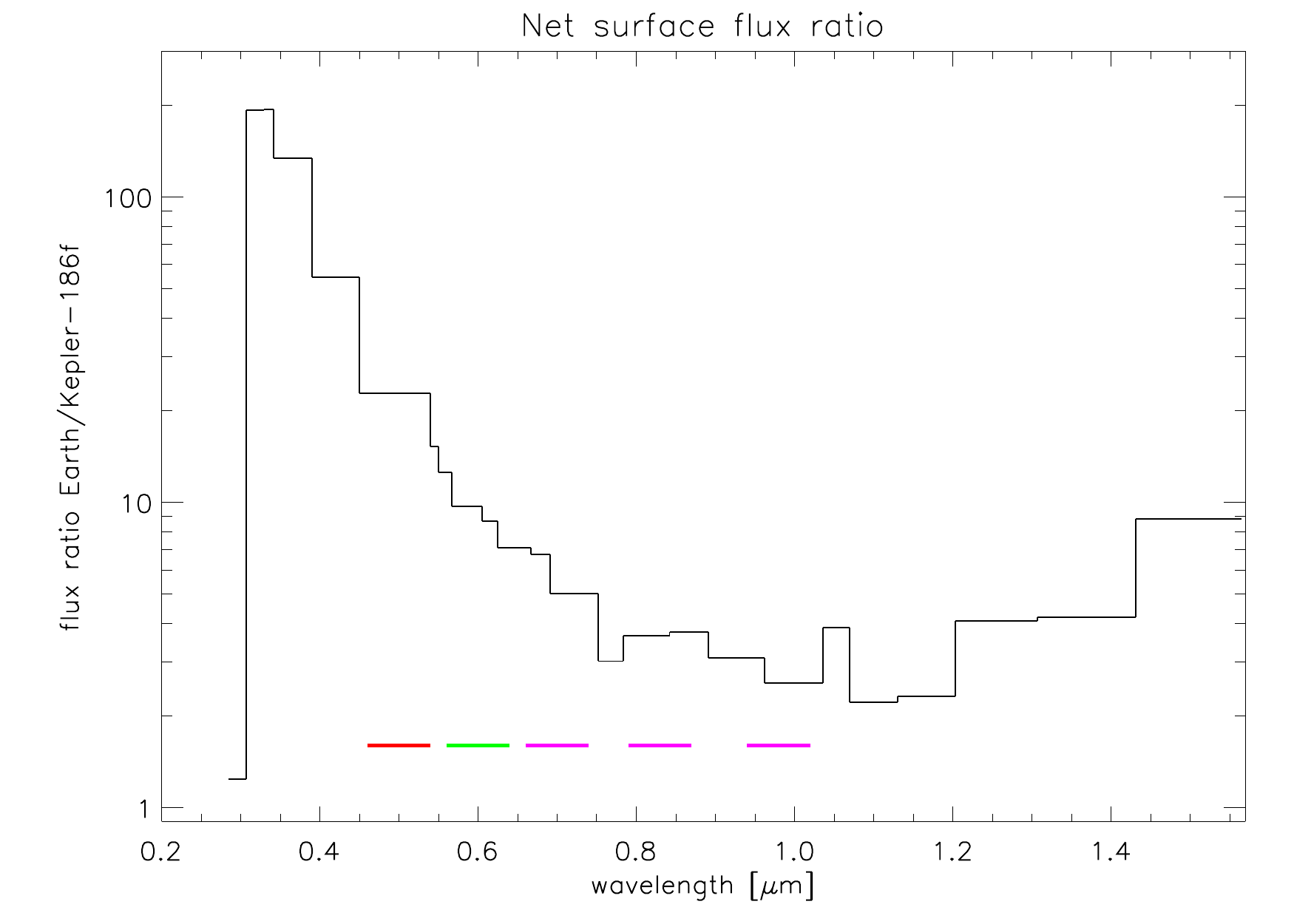}\\
  \caption{Ratio of net stellar flux at the surface between modern Earth and Set $\mathcal{B}$ (p$_{CO2}$=5\,bar, p$_{N2}$ =1\,bar, $g$=11.8\,ms$^{-2}$). Color bars indicate positions of photosynthetic pigments used by terrestrial biota.}\label{photo}
\end{figure}

A further interesting field of investigation would be the possibility of photosynthesis occurring on \starname f. A number of previous studies have investigated the potential for (an)oxygenic photosynthesis on planets orbiting M stars \citep[e.g.,][]{heath1999,kiang2007}. They found that photosynthesis is indeed possible, however, yielding less net productivity when assuming Earth-like pigment efficiencies. Figure \ref{photo} shows the ratio of the net surface radiative flux as a function of wavelength between modern Earth and a specific model scenario (p$_{CO2}$=5\,bar, p$_{N2}$ =1\,bar, $g$=11.8\,ms$^{-2}$) of Set $\mathcal{B}$. The surface temperature for this specific case is 285\,K, which is close to the surface temperature of modern Earth. Also indicated are the positions of photosynthetic pigments used by terrestrial biota. It is clearly seen that Earth's surface receives much more net radiative energy than the surface of \starname f (about a factor of six for the integrated flux).  At wavelengths around 500--700\,nm (corresponding to plant chlorophyll), the difference is even more pronounced  with modern Earth receiving 5--20 times more flux than \starname f. This suggests that even if \starname f is indeed habitable and life emerged, it is likely less productive than on Earth, in accordance with previous work \citep[e.g.,][]{heath1999,kiang2007}. This, at first glance, would imply that the detection of biosignatures is probably more difficult than for Earth-analogs. However, note that photosynthesis on Earth uses only a small fraction of the actually available sunlight before saturating \citep[a few to a few tens of percent, e.g,][]{heath1999}. Therefore, if the photosynthetic efficiency or the saturation threshold were higher, then productivity could be comparable to Earth. Note also that near-IR photosynthesis might be preferable on planets orbiting M stars instead of using visible wavelengths \citep{kiang2007}. Furthermore, an additional possibility to overcome the apparent lack of radiation is the development of multiple photosystems \citep{kiang2007}.

\begin{table*}[h]
  \centering
  \caption{Atmospheric Model Input Parameters.}\label{input_data}
  \begin{tabular}{lcl}
     \hline
   \hline
   Parameter & Value &Comment\\
    \hline
    \hline
     Insolation                            &   0.27--0.32 $S_{\oplus}$ &  \citet{Quintana2014} and Table \ref{tab:star} \\
    Gravity                            &   9.5--14.3\,ms$^{-2}$ & Range calculated from Table \ref{tab:planet} and Table \ref{tab:plcomp} \\
    $p_{{\rm CO2}}$                            &   0.1--10\,bar & Low estimates for Earth inventory \\
    $p_{{\rm N2}}$                            &   0--10\,bar & Uncertainty range for Earth \\
    Surface albedo               &   0.13 & Modern Earth value, \citet{rossow1999} \\
    Relative humidity               &  \citet{manabewetherald1967} & Used by  \citet{vparis2010gliese} and \citet{wordsworth2010}\\
 
     \hline
          \end{tabular}
\end{table*}

The orbits of the four inner planets remain interior to the inner boundary of the optimistic model.  However, it remains possible that Kepler-186e at least could have liquid water on its surface. As has been shown above, the rotation of Kepler-186e is most likely synchronized with its orbital period. Recent 3D model studies \citep[e.g.,][]{yang2013,yang2014} suggest that tidally locked or slowly rotating planets could remain habitable (albeit probably not in an Earth-centric sense) much closer to the central star than the traditional HZ limits. In addition, if Kepler-186e has significant cloud cover, its albedo is increased and the inner edge of the HZ can be pushed far inward of the optimistic criterion, into $\sim 0.5$ AU for a Sun-like star \citep{Selsis2007}.  

However, we point out that an orbit within the HZ does not make a planet habitable. Many other factors are required for a planet to be considered capable of hosting life \citep[e.g., available nutrients, magnetic fields, see e.g.][]{schulzemakuch2011}.  Even the list of factors for habitability is poorly constrained.  For example, the planet must have an adequate reservoir of volatiles, including water.  Given their rapid energetic accretion, water retention may in fact be a concern for planets orbiting low-mass stars \citep[e.g.,][]{Lissauer:2007,Raymond.etal:2007b}. It has also been proposed that plate tectonics is a key factor for maintaining a stable climate via the carbonate-silicate cycle \citep{walker1981}, however, plate tectonics does not occur on either Venus or on Mars. Plate tectonics may require a minimum internal heat flux from either the radioactive decay of long-lived isotopes \citep{williams1997moons} or tidal heating \citep{barnes2009}.  Venus' and Mars' orbits are both within certain estimates of the HZ but neither is thought to harbor life, partly because of their lack of plate tectonics, and, in the case of Mars, due to its small mass, and hence its inability to maintain a thick atmosphere. Since apart from its orbital period, radius and the central star, none of the habitability factors (atmosphere, bulk composition etc.) are known for \planetf (yet), our preliminary habitability assessment for \planetf is encouraging, but far from providing definitive conclusions.

In addition, it has been claimed that there is no real outer boundary to the HZ if atmospheric scenarios other than (broadly) Earth-like CO$_2$--N$_2$ cases are considered. The strong greenhouse effect of H$_2$-dominated atmospheres may possibly extend the HZ to an almost indefinite orbital distance \citep[e.g.,][]{stevenson1999,pierrehumbert2011hydro_hz,wordsworth2012}.

As has been shown in Section \ref{tidal} (see also, e.g., Figure \ref{fig:obl_f}), the rotation period  of planet f is likely of the order of days to weeks; even a complete 1:1 synchronization with the orbital period seems possible. If such synchronization did indeed happen, then the planet would possess permanent day- and nightsides. However, many previous studies have shown that slowly rotating planets likely have small latitudinal and longitudinal temperature gradients given atmospheric pressures of at least a few tens of millibars \citep[e.g.,][]{joshi1997,joshi2003,spiegel2008,wordsworth2011}). Therefore, at least in terms of surface temperature and atmospheric collapse, the habitability of planet f is not hindered by its increasing rotation period. Another possible concern is the potentially very high obliquity for extended periods of time (Figure \ref{fig:obl_f}). However, modeling studies by, e.g.,  \citet{williams1997moons} or \citet{spiegel2009}, suggest that high-obliquity climates are not necessarily an impediment to habitability. Furthermore, because of the very efficient energy transport to be expected for slow rotators \citep[e.g.,][]{joshi1997,wordsworth2011}, uneven stellar irradiation caused by high obliquity values will not influence habitability much.

\section{Summary and conclusions}
\label{conclusion}

We have presented an extensive study of the formation, orbital dynamics, tidal evolution, and habitability of the \starname system. 

In Section 2, we presented a simple end-to-end analysis of the accretion of the system.  Using the planets' orbital configuration, we built two minimum-mass disks.  We then attempted to reproduce the system's orbital architecture.  We performed simulations of in situ accretion from these disks, which we interpret as having been shaped by a previous episode of orbital migration.  The mass--orbital radius distribution of our simulations provided a modestly good match to the real system, although neither set of simulations adequately matches both the four inner planets and the outer one.  The planets also tended to have inclinations that were too large to be consistent with five planets in transit.  

Perhaps most striking is that our accretion simulations systematically formed too many planets.  At least one, and often two, planets tend to form between the orbits of \planete and \planetf.  From our dynamical simulations (Section \ref{effectextraplanet}), we can infer that if such a planet exists, then it should be less massive than $1~\mearth$, otherwise its gravitational influence on \planetf would most likely prevent \planetf from transiting. 

Given that the system is probably older than a few gigayears, simulations of tidal evolution show that the four inner planets of the system are in pseudo-synchronous rotation (respectively, P$_{\rm rot}$ = $\sim$4, $\sim$7, $\sim$13, $\sim$22, and $\sim$130~days) with very low obliquities ($<1\deg$). However, in a few simulations, the obliquity of \planetd was excited to more than 10$\deg$ due to a brief but deep crossing of the 5:3 mean motion resonance between \planetc and \planetd. The competition between the excitation due to planet--planet gravitational interactions and tidal damping has the effect of stabilizing this relatively high obliquity on $\sim$10~Myr timescales. 

We showed that the eccentricities of the planets cannot be as high as the upper value given by \citet{Quintana2014}. The maximum possible initial eccentricities depend of course on the mass of the planets (through their compositions). A system with planets made of 100\% iron can be stable over 1~Myr for small eccentricities ($\lesssim 0.04$). A system with Earth-composition planets can be stable over 1~Myr with higher eccentricities ($\sim 0.07$) but this can lead to an excitation of the inclinations inconsistent with a transit configuration. A system with 100\% ice planets can be stable over 1~Myr with eccentricities as high as 0.1 for the four inner planets and 0.4 for \planetf. Constraining the mass of the planets would be invaluable information to further constrain the dynamics of this system.

We also showed that given the uncertainties on the age of the star as well as the uncertainties on the composition and tidal dissipation, the rotation state of \planetf is unconstrained. If the system is somewhat younger--1~Gyr old--or if the tidal dissipation of \planetf is lower than that of Earth's, then \planetf could still be in the process of pseudo-synchronization and its obliquity could be high. However, if the system is about 4~Gyr old or the tidal dissipation of \planetf is Earth-like, then \planetf would be pseudo-synchronized with a long rotation period ($\sim130$~days). The variety of spin states of \planetf should then be investigated by exoplanet-climate modelers. Our calculations show that the tidal flux generated when the obliquity of the planet is still high is not sufficient to influence the atmosphere of \planetf.

The dynamics of the system would be affected if, as predicted by the accretion simulations, an additional planet existed between \planete and \planetf. Without this additional planet, \planetf is relatively isolated from the inner system, so its eccentricity and obliquity oscillations have a very low amplitude. However, an extra planet allows angular momentum to transfer from the inner parts of the system to \planetf, causing higher amplitude oscillations of eccentricity and obliquity. 

In Section \ref{habi} we presented atmospheric model calculations which indicate that \planetf is indeed squarely situated in the HZ around \starname, with relatively modest amounts of CO$_2$ and N$_2$ required to support conditions conducive to surface liquid water.

\acknowledgements 

The authors thank Gabriel Tobie for useful discussion on planets dissipation and the anonymous referee for constructive comments.  

S.N.R. and F.H. are grateful to the Agence Nationale pour la Recherche via grant ANR-13-BS05-0003-02 (project MOJO). S.N.R.'s contribution was performed as part of the NASA Astrobiology Institute's Virtual Planetary Laboratory Lead Team, supported by NASA under Cooperative Agreement No. NNA13AA93A.

This research has been partly supported by the Helmholtz Association through the research alliance ``Planetary Evolution and Life".
This study has received financial support from the French State in the frame of the ``Investments for the future" Programme IdEx Bordeaux, reference ANR-10-IDEX-03-02. 

F.S. acknowledges support from the European Research Council (Starting Grant 209622: E$_3$ARTHs)

\bibliographystyle{apj}

\begin{thebibliography}{}
\expandafter\ifx\csname natexlab\endcsname\relax\def\natexlab#1{#1}\fi

\bibitem[{{Andrews} \& {Williams}(2007{\natexlab{a}})}]{Andrews.etal:2007b}
{Andrews}, S.~M., \& {Williams}, J.~P. 2007{\natexlab{a}}, \apj, 671, 1800

\bibitem[{{Andrews} \& {Williams}(2007{\natexlab{b}})}]{Andrews.etal:2007a}
---. 2007{\natexlab{b}}, \apj, 659, 705

\bibitem[{{Armstrong} {et~al.}(2014){Armstrong}, {Barnes}, {Domagal-Goldman},
  {Breiner}, {Quinn}, \& {Meadows}}]{Armstrong2014}
{Armstrong}, J.~C., {Barnes}, R., {Domagal-Goldman}, S., {et~al.} 2014,
  \astrb, 14, 277

\bibitem[{{Barclay} {et~al.}(2013){Barclay}, {Burke}, {Howell}, {Rowe},
  {Huber}, {Isaacson}, {Jenkins}, {Kolbl}, {Marcy}, {Quintana}, {Still},
  {Twicken}, {Bryson}, {Borucki}, {Caldwell}, {Ciardi}, {Clarke},
  {Christiansen}, {Coughlin}, {Fischer}, {Li}, {Haas}, {Hunter}, {Lissauer},
  {Mullally}, {Sabale}, {Seader}, {Smith}, {Tenenbaum}, {Kamal Uddin}, \&
  {Thompson}}]{Barclay2013a}
{Barclay}, T., {Burke}, C.~J., {Howell}, S.~B., {et~al.} 2013, \apj, 768, 101

\bibitem[{{Barnes} {et~al.}(2009){Barnes}, {Jackson}, {Greenberg}, \&
  {Raymond}}]{barnes2009}
{Barnes}, R., {Jackson}, B., {Greenberg}, R., \& {Raymond}, S.~N. 2009, \apjl,
  700, L30

\bibitem[{{Barnes} {et~al.}(2013){Barnes}, {Mullins}, {Goldblatt}, {Meadows},
  {Kasting}, \& {Heller}}]{Barnes2013}
{Barnes}, R., {Mullins}, K., {Goldblatt}, C., {et~al.} 2013, \astrb, 13,
  225

\bibitem[{{Barnes} \& {Raymond}(2004)}]{barnes04}
{Barnes}, R., \& {Raymond}, S.~N. 2004, \apj, 617, 569

\bibitem[{{Batalha} {et~al.}(2011){Batalha}, {Borucki}, {Bryson}, {Buchhave},
  {Caldwell}, {Christensen-Dalsgaard}, {Ciardi}, {Dunham}, {Fressin},
  {Gautier}, {Gilliland}, {Haas}, {Howell}, {Jenkins}, {Kjeldsen}, {Koch},
  {Latham}, {Lissauer}, {Marcy}, {Rowe}, {Sasselov}, {Seager}, {Steffen},
  {Torres}, {Basri}, {Brown}, {Charbonneau}, {Christiansen}, {Clarke},
  {Cochran}, {Dupree}, {Fabrycky}, {Fischer}, {Ford}, {Fortney}, {Girouard},
  {Holman}, {Johnson}, {Isaacson}, {Klaus}, {Machalek}, {Moorehead},
  {Morehead}, {Ragozzine}, {Tenenbaum}, {Twicken}, {Quinn}, {VanCleve},
  {Walkowicz}, {Welsh}, {Devore}, \& {Gould}}]{Batalha.etal.:2011}
{Batalha}, N.~M., {Borucki}, W.~J., {Bryson}, S.~T., {et~al.} 2011, \apj, 729,
  27

\bibitem[{{Batygin} {et~al.}(2009){Batygin}, {Bodenheimer}, \&
  {Laughlin}}]{Batygin2009}
{Batygin}, K., {Bodenheimer}, P., \& {Laughlin}, G. 2009, \apjl, 704, L49

\bibitem[{{Berger}(1988)}]{berger1988}
{Berger}, A. 1988, Reviews of Geophysics, 26, 624

\bibitem[{{Bitsch} {et~al.}(2013){Bitsch}, {Crida}, {Morbidelli}, {Kley}, \&
  {Dobbs-Dixon}}]{bitsch13}
{Bitsch}, B., {Crida}, A., {Morbidelli}, A., {Kley}, W., \& {Dobbs-Dixon}, I.
  2013, \aap, 549, A124

\bibitem[{{Boley} \& {Ford}(2013)}]{boley13}
{Boley}, A.~C., \& {Ford}, E.~B. 2013, arXiv:1306.0566

\bibitem[{{Bolmont} {et~al.}(2013){Bolmont}, {Selsis}, {Raymond}, {Leconte},
  {Hersant}, {Maurin}, \& {Pericaud}}]{Bolmont.etal:2013}
{Bolmont}, E., {Selsis}, F., {Raymond}, S.~N., {et~al.} 2013, \aap, 556, A17

\bibitem[{{Borucki} {et~al.}(2010){Borucki}, {Koch}, {Basri}, {Batalha},
  {Brown}, {Caldwell}, {Caldwell}, {Christensen-Dalsgaard}, {Cochran},
  {DeVore}, {Dunham}, {Dupree}, {Gautier}, {Geary}, {Gilliland}, {Gould},
  {Howell}, {Jenkins}, {Kondo}, {Latham}, {Marcy}, {Meibom}, {Kjeldsen},
  {Lissauer}, {Monet}, {Morrison}, {Sasselov}, {Tarter}, {Boss}, {Brownlee},
  {Owen}, {Buzasi}, {Charbonneau}, {Doyle}, {Fortney}, {Ford}, {Holman},
  {Seager}, {Steffen}, {Welsh}, {Rowe}, {Anderson}, {Buchhave}, {Ciardi},
  {Walkowicz}, {Sherry}, {Horch}, {Isaacson}, {Everett}, {Fischer}, {Torres},
  {Johnson}, {Endl}, {MacQueen}, {Bryson}, {Dotson}, {Haas}, {Kolodziejczak},
  {Van Cleve}, {Chandrasekaran}, {Twicken}, {Quintana}, {Clarke}, {Allen},
  {Li}, {Wu}, {Tenenbaum}, {Verner}, {Bruhweiler}, {Barnes}, \&
  {Prsa}}]{borucki2010}
{Borucki}, W.~J., {Koch}, D., {Basri}, G., {et~al.} 2010, Science, 327, 977

\bibitem[{{Borucki} {et~al.}(2011){Borucki}, {Koch}, {Basri}, {Batalha},
  {Brown}, {Bryson}, {Caldwell}, {Christensen-Dalsgaard}, {Cochran}, {DeVore},
  {Dunham}, {Gautier}, {Geary}, {Gilliland}, {Gould}, {Howell}, {Jenkins},
  {Latham}, {Lissauer}, {Marcy}, {Rowe}, {Sasselov}, {Boss}, {Charbonneau},
  {Ciardi}, {Doyle}, {Dupree}, {Ford}, {Fortney}, {Holman}, {Seager},
  {Steffen}, {Tarter}, {Welsh}, {Allen}, {Buchhave}, {Christiansen}, {Clarke},
  {Das}, {D{\'e}sert}, {Endl}, {Fabrycky}, {Fressin}, {Haas}, {Horch},
  {Howard}, {Isaacson}, {Kjeldsen}, {Kolodziejczak}, {Kulesa}, {Li}, {Lucas},
  {Machalek}, {McCarthy}, {MacQueen}, {Meibom}, {Miquel}, {Prsa}, {Quinn},
  {Quintana}, {Ragozzine}, {Sherry}, {Shporer}, {Tenenbaum}, {Torres},
  {Twicken}, {Van Cleve}, {Walkowicz}, {Witteborn}, \& {Still}}]{borucki2011}
{Borucki}, W.~J., {Koch}, D.~G., {Basri}, G., {et~al.} 2011, \apj, 736, 19

\bibitem[{{Borucki} {et~al.}(2012){Borucki}, {Koch}, {Batalha}, {Bryson},
  {Rowe}, {Fressin}, {Torres}, {Caldwell}, {Christensen-Dalsgaard}, {Cochran},
  {DeVore}, {Gautier}, {Geary}, {Gilliland}, {Gould}, {Howell}, {Jenkins},
  {Latham}, {Lissauer}, {Marcy}, {Sasselov}, {Boss}, {Charbonneau}, {Ciardi},
  {Kaltenegger}, {Doyle}, {Dupree}, {Ford}, {Fortney}, {Holman}, {Steffen},
  {Mullally}, {Still}, {Tarter}, {Ballard}, {Buchhave}, {Carter},
  {Christiansen}, {Demory}, {D{\'e}sert}, {Dressing}, {Endl}, {Fabrycky},
  {Fischer}, {Haas}, {Henze}, {Horch}, {Howard}, {Isaacson}, {Kjeldsen},
  {Johnson}, {Klaus}, {Kolodziejczak}, {Barclay}, {Li}, {Meibom}, {Prsa},
  {Quinn}, {Quintana}, {Robertson}, {Sherry}, {Shporer}, {Tenenbaum},
  {Thompson}, {Twicken}, {Van Cleve}, {Welsh}, {Basu}, {Chaplin}, {Miglio},
  {Kawaler}, {Arentoft}, {Stello}, {Metcalfe}, {Verner}, {Karoff}, {Lundkvist},
  {Lund}, {Handberg}, {Elsworth}, {Hekker}, {Huber}, {Bedding}, \&
  {Rapin}}]{Borucki.etal:2012}
{Borucki}, W.~J., {Koch}, D.~G., {Batalha}, N., {et~al.} 2012, \apj, 745, 120

\bibitem[{{Borucki} {et~al.}(2013){Borucki}, {Agol}, {Fressin}, {Kaltenegger},
  {Rowe}, {Isaacson}, {Fischer}, {Batalha}, {Lissauer}, {Marcy}, {Fabrycky},
  {D{\'e}sert}, {Bryson}, {Barclay}, {Bastien}, {Boss}, {Brugamyer},
  {Buchhave}, {Burke}, {Caldwell}, {Carter}, {Charbonneau}, {Crepp},
  {Christensen-Dalsgaard}, {Christiansen}, {Ciardi}, {Cochran}, {DeVore},
  {Doyle}, {Dupree}, {Endl}, {Everett}, {Ford}, {Fortney}, {Gautier}, {Geary},
  {Gould}, {Haas}, {Henze}, {Howard}, {Howell}, {Huber}, {Jenkins}, {Kjeldsen},
  {Kolbl}, {Kolodziejczak}, {Latham}, {Lee}, {Lopez}, {Mullally}, {Orosz},
  {Prsa}, {Quintana}, {Sanchis-Ojeda}, {Sasselov}, {Seader}, {Shporer},
  {Steffen}, {Still}, {Tenenbaum}, {Thompson}, {Torres}, {Twicken}, {Welsh}, \&
  {Winn}}]{Borucki.etal:2013}
{Borucki}, W.~J., {Agol}, E., {Fressin}, F., {et~al.} 2013, Science, 340, 587

\bibitem[{{Chambers}(1999)}]{Chambers:1999}
{Chambers}, J.~E. 1999, \mnras, 304, 793

\bibitem[{{Chambers} {et~al.}(1996){Chambers}, {Wetherill}, \&
  {Boss}}]{Chambers.etal:1996}
{Chambers}, J.~E., {Wetherill}, G.~W., \& {Boss}, A.~P. 1996, Icarus, 119, 261

\bibitem[{{Chatterjee} \& {Tan}(2014)}]{chatterjee14}
{Chatterjee}, S., \& {Tan}, J.~C. 2014, \apj, 780, 53

\bibitem[{{Chiang} \& {Laughlin}(2013)}]{Chiang.etal:2013}
{Chiang}, E., \& {Laughlin}, G. 2013, \mnras, 431, 3444

\bibitem[{{Correia} \& {Rodr{\'{\i}}guez}(2013)}]{CorreiaRodriguez2013}
{Correia}, A.~C.~M., \& {Rodr{\'{\i}}guez}, A. 2013, \apj, 767, 128

\bibitem[{{Cossou} {et~al.}(2014){Cossou}, {Raymond}, \& {Pierens}}]{cossou13b}
{Cossou}, C., {Raymond}, S.~N., \& {Pierens}, A. 2014, IAU Symposium 299, Exploring the formation and evolution of planetary systems, 360--364

\bibitem[{{Dobbs-Dixon} {et~al.}(2004){Dobbs-Dixon}, {Lin}, \&
  {Mardling}}]{DobbsDixon2004}
{Dobbs-Dixon}, I., {Lin}, D.~N.~C., \& {Mardling}, R.~A. 2004, \apj, 610, 464

\bibitem[{{Dole}(1964)}]{dole1964}
{Dole}, S.~H. 1964, {Habitable planets for man}

\bibitem[{{Fang} \& {Margot}(2012)}]{Fang2012}
{Fang}, J., \& {Margot}, J.-L. 2012, \apj, 761, 92

\bibitem[{{Ferraz-Mello} {et~al.}(2008){Ferraz-Mello}, {Rodr{\'{\i}}guez}, \&
  {Hussmann}}]{Ferrazmello.etal:2008}
{Ferraz-Mello}, S., {Rodr{\'{\i}}guez}, A., \& {Hussmann}, H. 2008, Celestial
  Mechanics and Dynamical Astronomy, 101, 171

\bibitem[{{Fortney} {et~al.}(2007){Fortney}, {Marley}, \&
  {Barnes}}]{Fortney.etal:2007}
{Fortney}, J.~J., {Marley}, M.~S., \& {Barnes}, J.~W. 2007, \apj, 659, 1661

\bibitem[{{Fressin} {et~al.}(2012){Fressin}, {Torres}, {Rowe}, {Charbonneau},
  {Rogers}, {Ballard}, {Batalha}, {Borucki}, {Bryson}, {Buchhave}, {Ciardi},
  {D{\'e}sert}, {Dressing}, {Fabrycky}, {Ford}, {Gautier}, {Henze}, {Holman},
  {Howard}, {Howell}, {Jenkins}, {Koch}, {Latham}, {Lissauer}, {Marcy},
  {Quinn}, {Ragozzine}, {Sasselov}, {Seager}, {Barclay}, {Mullally}, {Seader},
  {Still}, {Twicken}, {Thompson}, \& {Uddin}}]{Fressin.etal.2012}
{Fressin}, F., {Torres}, G., {Rowe}, J.~F., {et~al.} 2012, \nat, 482, 195

\bibitem[{{Gladman}(1993)}]{Gladman1993}
{Gladman}, B. 1993, Icarus, 106, 247

\bibitem[{{Goldblatt} {et~al.}(2009){Goldblatt}, {Claire}, {Lenton},
  {Matthews}, {Watson}, \& {Zahnle}}]{goldblatt2009faintyoungsun}
{Goldblatt}, C., {Claire}, M.~W., {Lenton}, T.~M., {et~al.} 2009, Nature
  Geoscience, 2, 891, 896

\bibitem[{{Goldreich} \& {Tremaine}(1980)}]{goldreich80}
{Goldreich}, P., \& {Tremaine}, S. 1980, \apj, 241, 425

\bibitem[{{Grenfell} {et~al.}(2010){Grenfell}, {Rauer}, {Selsis},
  {Kaltenegger}, {Beichman}, {Danchi}, {Eiroa}, {Fridlund}, {Henning},
  {Herbst}, {Lammer}, {L{\'e}ger}, {Liseau}, {Lunine}, {Paresce}, {Penny},
  {Quirrenbach}, {R{\"o}ttgering}, {Schneider}, {Stam}, {Tinetti}, \&
  {White}}]{grenfell2010}
{Grenfell}, J.~L., {Rauer}, H., {Selsis}, F., {et~al.} 2010, \astrb, 10,
  77

\bibitem[{{Haisch} {et~al.}(2001){Haisch}, {Lada}, \&
  {Lada}}]{Haisch.etal:2001}
{Haisch}, Jr., K.~E., {Lada}, E.~A., \& {Lada}, C.~J. 2001, \apjl, 553, L153

\bibitem[{{Hansen} \& {Murray}(2012)}]{Hansen.etal:2012}
{Hansen}, B.~M.~S., \& {Murray}, N. 2012, \apj, 751, 158

\bibitem[{{Hansen} \& {Murray}(2013)}]{Hansen2013}
---. 2013, \apj, 775, 53

\bibitem[{{Hart}(1978)}]{hart1978}
{Hart}, M.~H. 1978, \icarus, 33, 23

\bibitem[{{Hauschildt} {et~al.}(1999){Hauschildt}, {Allard}, \&
  {Baron}}]{hauschildt1999}
{Hauschildt}, P.~H., {Allard}, F., \& {Baron}, E. 1999, \apj, 512, 377

\bibitem[{{Heath} {et~al.}(1999){Heath}, {Doyle}, {Joshi}, \&
  {Haberle}}]{heath1999}
{Heath}, M.~J., {Doyle}, L.~R., {Joshi}, M.~M., \& {Haberle}, R.~M. 1999,
  Origins of Life and Evolution of the Biosphere, 29, 405

\bibitem[{{Heller} {et~al.}(2011){Heller}, {Leconte}, \& {Barnes}}]{heller2011}
{Heller}, R., {Leconte}, J., \& {Barnes}, R. 2011, \aap, 528, A27

\bibitem[{{Hillenbrand}(2008)}]{Hillenbrand:2008}
{Hillenbrand}, L.~A. 2008, Physica Scripta Volume T, 130, 014024

\bibitem[{{Hirano} {et~al.}(2012){Hirano}, {Narita}, {Sato}, {Takahashi},
  {Masuda}, {Takeda}, {Aoki}, {Tamura}, \& {Suto}}]{hirano12}
{Hirano}, T., {Narita}, N., {Sato}, B., {et~al.} 2012, \apjl, 759, L36

\bibitem[{{Hu} \& {Ding}(2011)}]{hu2011}
{Hu}, Y., \& {Ding}, F. 2011, \aa, 526, A135

\bibitem[{{Hut}(1981)}]{Hut:1981}
{Hut}, P. 1981, \aap, 99, 126

\bibitem[{{Jackson} {et~al.}(2000){Jackson}, {Fitz Gerald}, \&
  Kokkonen}]{JacksonFitzGerald2000}
{Jackson}, I., {Fitz Gerald}, J.~D., \& Kokkonen, H. 2000, J. Geophys. Res.,
  105, 23605

\bibitem[{{Jacobson} \& {Morbidelli}(2014)}]{jacobson14}
{Jacobson}, S.~A., \& {Morbidelli}, A. 2014, arXiv:1406.2697

\bibitem[{{Jones} {et~al.}(2006){Jones}, {Sleep}, \& {Underwood}}]{jones2006}
{Jones}, B.~W., {Sleep}, P.~N., \& {Underwood}, D.~R. 2006, \apj, 649, 1010

\bibitem[{{Jontof-Hutter} {et~al.}(2014){Jontof-Hutter}, {Lissauer}, {Rowe}, \&
  {Fabrycky}}]{jontofhutter14}
{Jontof-Hutter}, D., {Lissauer}, J.~J., {Rowe}, J.~F., \& {Fabrycky}, D.~C.
  2014, \apj, 785, 15

\bibitem[{{Joshi}(2003)}]{joshi2003}
{Joshi}, M. 2003, \astrb, 3, 415

\bibitem[{{Joshi} {et~al.}(1997){Joshi}, {Haberle}, \& {Reynolds}}]{joshi1997}
{Joshi}, M.~M., {Haberle}, R.~M., \& {Reynolds}, R.~T. 1997, \icarus, 129, 450

\bibitem[{{Kaltenegger} {et~al.}(2011){Kaltenegger}, {Segura}, \&
  {Mohanty}}]{kaltenegger2011}
{Kaltenegger}, L., {Segura}, A., \& {Mohanty}, S. 2011, \apj, 733, 35

\bibitem[{{Kasting}(1988)}]{kasting1988}
{Kasting}, J.~F. 1988, \icarus, 74, 472

\bibitem[{{Kasting} {et~al.}(1993){Kasting}, {Whitmire}, \&
  {Reynolds}}]{kasting1993}
{Kasting}, J.~F., {Whitmire}, D.~P., \& {Reynolds}, R.~T. 1993, \icarus, 101,
  108

\bibitem[{{Kennedy} \& {Kenyon}(2008)}]{kennedy08}
{Kennedy}, G.~M., \& {Kenyon}, S.~J. 2008, \apj, 673, 502

\bibitem[{{Kiang} {et~al.}(2007){Kiang}, {Segura}, {Tinetti}, {Govindjee},
  {Blankenship}, {Cohen}, {Siefert}, {Crisp}, \& {Meadows}}]{kiang2007}
{Kiang}, N.~Y., {Segura}, A., {Tinetti}, G., {et~al.} 2007, \astrb, 7,
  252

\bibitem[{{Kidder}(1995)}]{kidder1995}
{Kidder}, L.~E. 1995, \prd, 52, 821

\bibitem[{{Kokubo} \& {Ida}(2007)}]{kokubo07}
{Kokubo}, E., \& {Ida}, S. 2007, \apj, 671, 2082

\bibitem[{{Kokubo} {et~al.}(2006{\natexlab{a}}){Kokubo}, {Kominami}, \&
  {Ida}}]{Kokubo.etal:2006}
{Kokubo}, E., {Kominami}, J., \& {Ida}, S. 2006{\natexlab{a}}, \apj, 642, 1131

\bibitem[{{Koot} \& {Dumberry}(2011)}]{KootDumberry2011}
{Koot}, L., \& {Dumberry}, M. 2011, Earth and Planetary Science Letters, 308,
  343

\bibitem[{{Kopparapu} {et~al.}(2013){Kopparapu}, {Ramirez}, {Kasting}, {Eymet},
  {Robinson}, {Mahadevan}, {Terrien}, {Domagal-Goldman}, {Meadows}, \&
  {Deshpande}}]{kopparapu2013}
{Kopparapu}, R., {Ramirez}, R., {Kasting}, J., {et~al.} 2013, \apj, 765, 131

\bibitem[{{Kopparapu} \& {Barnes}(2010)}]{kopparapu2010}
{Kopparapu}, R.~K., \& {Barnes}, R. 2010, \apj, 716, 1336

\bibitem[{{Kopparapu} {et~al.}(2014){Kopparapu}, {Ramirez}, {SchottelKotte},
  {Kasting}, {Domagal-Goldman}, \& {Eymet}}]{kopparapu2014}
{Kopparapu}, R.~K., {Ramirez}, R.~M., {SchottelKotte}, J., {et~al.} 2014,
  \apjl, 787, L29

\bibitem[{{Lambeck}(1977)}]{Lambeck1977}
{Lambeck}, K. 1977, Royal Society of London Philosophical Transactions Series
  A, 287, 545

\bibitem[{{Lammer} {et~al.}(2010){Lammer}, {Selsis}, {Chassefi{\`e}re},
  {Breuer}, {Grie{\ss}meier}, {Kulikov}, {Erkaev}, {Khodachenko}, {Biernat},
  {Leblanc}, {Kallio}, {Lundin}, {Westall}, {Bauer}, {Beichman}, {Danchi},
  {Eiroa}, {Fridlund}, {Gr{\"o}ller}, {Hanslmeier}, {Hausleitner}, {Henning},
  {Herbst}, {Kaltenegger}, {L{\'e}ger}, {Leitzinger}, {Lichtenegger}, {Liseau},
  {Lunine}, {Motschmann}, {Odert}, {Paresce}, {Parnell}, {Penny},
  {Quirrenbach}, {Rauer}, {R{\"o}ttgering}, {Schneider}, {Spohn}, {Stadelmann},
  {Stangl}, {Stam}, {Tinetti}, \& {White}}]{lammer2010}
{Lammer}, H., {Selsis}, F., {Chassefi{\`e}re}, E., {et~al.} 2010, \astrb,
  10, 45

\bibitem[{{Leconte} {et~al.}(2010){Leconte}, {Chabrier}, {Baraffe}, \&
  {Levrard}}]{leconte2010}
{Leconte}, J., {Chabrier}, G., {Baraffe}, I., \& {Levrard}, B. 2010, \aap, 516,
  A64

\bibitem[{{Li} {et~al.}(2009){Li}, {Pahlevan}, {Kirschvink}, \&
  {Yung}}]{Li2009}
{Li}, K., {Pahlevan}, K., {Kirschvink}, J., \& {Yung}, Y. 2009, Proceedings of
  the National Academy of Sciences, 106, 9576

\bibitem[{{Lissauer}(2007)}]{Lissauer:2007}
{Lissauer}, J.~J. 2007, \apjl, 660, L149

\bibitem[{{Lissauer} {et~al.}(2011){Lissauer}, {Ragozzine}, {Fabrycky},
  {Steffen}, {Ford}, {Jenkins}, {Shporer}, {Holman}, {Rowe}, {Quintana},
  {Batalha}, {Borucki}, {Bryson}, {Caldwell}, {Carter}, {Ciardi}, {Dunham},
  {Fortney}, {Gautier}, {Howell}, {Koch}, {Latham}, {Marcy}, {Morehead}, \&
  {Sasselov}}]{Lissauer.etal:2011}
{Lissauer}, J.~J., {Ragozzine}, D., {Fabrycky}, D.~C., {et~al.} 2011, \apjs,
  197, 8

\bibitem[{{Lopez} \& {Fortney}(2013)}]{lopez13}
{Lopez}, E.~D., \& {Fortney}, J.~J. 2013, arXiv:1311.0329

\bibitem[{{Lundin} \& {Barabash}(2004)}]{lundin2004}
{Lundin}, R., \& {Barabash}, S. 2004, \planss, 52, 1059

\bibitem[{{Makarov} \& {Efroimsky}(2013)}]{Makarov.etal:2013}
{Makarov}, V.~V., \& {Efroimsky}, M. 2013, \apj, 764, 27

\bibitem[{{Manabe} \& {Wetherald}(1967)}]{manabewetherald1967}
{Manabe}, S., \& {Wetherald}, R.~T. 1967, \jas, 24, 241

\bibitem[{{Marchal} \& {Bozis}(1982)}]{Marchal.etal:1982}
{Marchal}, C., \& {Bozis}, G. 1982, Celestial Mechanics, 26, 311

\bibitem[{{Marcy} {et~al.}(2014{\natexlab{a}}){Marcy}, {Weiss}, {Petigura},
  {Isaacson}, {Howard}, \& {Buchhave}}]{marcy2014b}
{Marcy}, G.~W., {Weiss}, L.~M., {Petigura}, E.~A., {et~al.} 2014{\natexlab{a}},
  arXiv:1404.2960

\bibitem[{{Marcy} {et~al.}(2014{\natexlab{b}}){Marcy}, {Isaacson}, {Howard},
  {Rowe}, {Jenkins}, {Bryson}, {Latham}, {Howell}, {Gautier}, {Batalha},
  {Rogers}, {Ciardi}, {Fischer}, {Gilliland}, {Kjeldsen},
  {Christensen-Dalsgaard}, {Huber}, {Chaplin}, {Basu}, {Buchhave}, {Quinn},
  {Borucki}, {Koch}, {Hunter}, {Caldwell}, {Van Cleve}, {Kolbl}, {Weiss},
  {Petigura}, {Seager}, {Morton}, {Johnson}, {Ballard}, {Burke}, {Cochran},
  {Endl}, {MacQueen}, {Everett}, {Lissauer}, {Ford}, {Torres}, {Fressin},
  {Brown}, {Steffen}, {Charbonneau}, {Basri}, {Sasselov}, {Winn},
  {Sanchis-Ojeda}, {Christiansen}, {Adams}, {Henze}, {Dupree}, {Fabrycky},
  {Fortney}, {Tarter}, {Holman}, {Tenenbaum}, {Shporer}, {Lucas}, {Welsh},
  {Orosz}, {Bedding}, {Campante}, {Davies}, {Elsworth}, {Handberg}, {Hekker},
  {Karoff}, {Kawaler}, {Lund}, {Lundkvist}, {Metcalfe}, {Miglio}, {Silva
  Aguirre}, {Stello}, {White}, {Boss}, {Devore}, {Gould}, {Prsa}, {Agol},
  {Barclay}, {Coughlin}, {Brugamyer}, {Mullally}, {Quintana}, {Still},
  {Thompson}, {Morrison}, {Twicken}, {D{\'e}sert}, {Carter}, {Crepp},
  {H{\'e}brard}, {Santerne}, {Moutou}, {Sobeck}, {Hudgins}, {Haas},
  {Robertson}, {Lillo-Box}, \& {Barrado}}]{Marcy2014}
{Marcy}, G.~W., {Isaacson}, H., {Howard}, A.~W., {et~al.} 2014{\natexlab{b}},
  \apjs, 210, 20

\bibitem[{{Mardling}(2007)}]{Mardling:2007}
{Mardling}, R.~A. 2007, \mnras, 382, 1768

\bibitem[{{Marzari} {et~al.}(2002){Marzari}, {Tricarico}, \&
  {Scholl}}]{Marzari2002}
{Marzari}, F., {Tricarico}, P., \& {Scholl}, H. 2002, \apj, 579, 905

\bibitem[{{Masset} {et~al.}(2006){Masset}, {Morbidelli}, {Crida}, \&
  {Ferreira}}]{masset06}
{Masset}, F.~S., {Morbidelli}, A., {Crida}, A., \& {Ferreira}, J. 2006, \apj,
  642, 478

\bibitem[{{Mayor} {et~al.}(2009){Mayor}, {Bonfils}, {Forveille}, {Delfosse},
  {Udry}, {Bertaux}, {Beust}, {Bouchy}, {Lovis}, {Pepe}, {Perrier}, {Queloz},
  \& {Santos}}]{mayor09}
{Mayor}, M., {Bonfils}, X., {Forveille}, T., {et~al.} 2009, \aap, 507, 487

\bibitem[{{McCarthy} \& {Castillo-Rogez}(2013)}]{McCarthyCastillo2013}
{McCarthy}, C., \& {Castillo-Rogez}, J.~C. 2013, {Planetary Ices Attenuation
  Properties}, ed. M.~S. {Gudipati} \& J.~{Castillo-Rogez}, Springer, New York 183

\bibitem[{{McKay} \& {Stoker}(1989)}]{mckay1989mars}
{McKay}, C.~P., \& {Stoker}, C.~R. 1989, Reviews of Geophysics, 27, 189

\bibitem[{{Menou} \& {Tabachnik}(2003)}]{menou03}
{Menou}, K., \& {Tabachnik}, S. 2003, \apj, 583, 473

\bibitem[{{Morbidelli} {et~al.}(2000){Morbidelli}, {Chambers}, {Lunine},
  {Petit}, {Robert}, {Valsecchi}, \& {Cyr}}]{morby00}
{Morbidelli}, A., {Chambers}, J., {Lunine}, J.~I., {et~al.} 2000, Meteoritics
  and Planetary Science, 35, 1309

\bibitem[{{Mundy} {et~al.}(2000){Mundy}, {Looney}, \&
  {Welch}}]{Mundy.etal:2000}
{Mundy}, L.~G., {Looney}, L.~W., \& {Welch}, W.~J. 2000, Protostars and Planets
  IV, University of Arizona Press, Tucson, 355

\bibitem[{{Murray} \& {Dermott}(1999)}]{Murray.etal:1999}
{Murray}, C.~D., \& {Dermott}, S.~F. 1999, {Solar system dynamics}, Princeton

\bibitem[{{Neron de Surgy} \& {Laskar}(1997)}]{DeSurgyLaskar1997}
{Neron de Surgy}, O., \& {Laskar}, J. 1997, A \& A, 318, 975

\bibitem[{{O'Brien} {et~al.}(2006){O'Brien}, {Morbidelli}, \&
  {Levison}}]{obrien06}
{O'Brien}, D.~P., {Morbidelli}, A., \& {Levison}, H.~F. 2006, \icarus, 184, 39

\bibitem[{{O'Brien} {et~al.}(2014){O'Brien}, {Walsh}, {Morbidelli}, {Raymond},
  \& {Mandell}}]{obrien14}
{O'Brien}, D.~P., {Walsh}, K.~J., {Morbidelli}, A., {Raymond}, S.~N., \&
  {Mandell}, A.~M. 2014, \icarus, 239, 74

\bibitem[{{Ogihara} \& {Ida}(2009)}]{Ogihara.etal:2009}
{Ogihara}, M., \& {Ida}, S. 2009, \apj, 699, 824

\bibitem[{{Ohmoto} {et~al.}(2004){Ohmoto}, {Watanabe}, \&
  {Kumazawa}}]{ohmoto2004}
{Ohmoto}, H., {Watanabe}, Y., \& {Kumazawa}, K. 2004, \nat, 429, 395

\bibitem[{{Pierens} {et~al.}(2013){Pierens}, {Cossou}, \&
  {Raymond}}]{pierens13}
{Pierens}, A., {Cossou}, C., \& {Raymond}, S.~N. 2013, \aap, 558, A105

\bibitem[{{Pierrehumbert} \& {Gaidos}(2011)}]{pierrehumbert2011hydro_hz}
{Pierrehumbert}, R., \& {Gaidos}, E. 2011, \apjl, 734, L13

\bibitem[{{Quintana} \& {Lissauer}(2014)}]{quintanalissauer2014}
{Quintana}, E.~V., \& {Lissauer}, J.~J. 2014, \apj, 786, 33

\bibitem[{{Quintana} {et~al.}(2014){Quintana}, {Barclay}, {Raymond}, {Rowe},
  {Bolmont}, {Caldwell}, {Howell}, {Kane}, {Huber}, {Crepp}, {Lissauer},
  {Ciardi}, {Coughlin}, {Everett}, {Henze}, {Horch}, {Isaacson}, {Ford},
  {Adams}, {Still}, {Hunter}, {Quarles}, \& {Selsis}}]{Quintana2014}
{Quintana}, E.~V., {Barclay}, T., {Raymond}, S.~N., {et~al.} 2014, Science,
  344, 277

\bibitem[{{Ragozzine} \& {Holman}(2010)}]{ragozzine10}
{Ragozzine}, D., \& {Holman}, M.~J. 2010, arXiv:1006.3727

\bibitem[{{Rauer} {et~al.}(2011){Rauer}, {Gebauer}, {von Paris}, {Cabrera},
  {Godolt}, {Grenfell}, {Belu}, {Selsis}, {Hedelt}, \& {Schreier}}]{rauer2011}
{Rauer}, H., {Gebauer}, S., {von Paris}, P., {et~al.} 2011, \aa, 529, A8

\bibitem[{{Raymond}(2006)}]{raymond06a}
{Raymond}, S.~N. 2006, \apjl, 643, L131

\bibitem[{{Raymond} {et~al.}(2008{\natexlab{a}}){Raymond}, {Barnes}, \&
  {Mandell}}]{raymond08}
{Raymond}, S.~N., {Barnes}, R., \& {Mandell}, A.~M. 2008{\natexlab{a}}, \mnras,
  384, 663

\bibitem[{{Raymond} \& {Cossou}(2014)}]{Raymond:2014}
{Raymond}, S.~N., \& {Cossou}, C. 2014, \mnras, 440, L11

\bibitem[{{Raymond} {et~al.}(2013){Raymond}, {Kokubo}, {Morbidelli},
  {Morishima}, \& {Walsh}}]{raymond14pp6}
{Raymond}, S.~N., {Kokubo}, E., {Morbidelli}, A., {Morishima}, R., \& {Walsh},
  K.~J. 2013, arXiv:1312.1689

\bibitem[{{Raymond} {et~al.}(2009){Raymond}, {O'Brien}, {Morbidelli}, \&
  {Kaib}}]{raymond09c}
{Raymond}, S.~N., {O'Brien}, D.~P., {Morbidelli}, A., \& {Kaib}, N.~A. 2009,
  Icarus, 203, 644

\bibitem[{{Raymond} {et~al.}(2004){Raymond}, {Quinn}, \& {Lunine}}]{raymond04}
{Raymond}, S.~N., {Quinn}, T., \& {Lunine}, J.~I. 2004, Icarus, 168, 1

\bibitem[{{Raymond} {et~al.}(2005{\natexlab{a}}){Raymond}, {Quinn}, \&
  {Lunine}}]{raymond05b}
---. 2005{\natexlab{a}}, \apj, 632, 670

\bibitem[{{Raymond} {et~al.}(2006){Raymond}, {Quinn}, \& {Lunine}}]{raymond06}
---. 2006, \icarus, 183, 265

\bibitem[{{Raymond} {et~al.}(2007{\natexlab{a}}){Raymond}, {Quinn}, \&
  {Lunine}}]{raymond07}
---. 2007{\natexlab{a}}, \astrb, 7, 66

\bibitem[{{Raymond} {et~al.}(2007{\natexlab{b}}){Raymond}, {Scalo}, \&
  {Meadows}}]{Raymond.etal:2007b}
{Raymond}, S.~N., {Scalo}, J., \& {Meadows}, V.~S. 2007{\natexlab{b}}, \apj,
  669, 606

\bibitem[{{Ronco} \& {de El{\'{\i}}a}(2014)}]{RoncoDeElia2014}
{Ronco}, M.~P., \& {de El{\'{\i}}a}, G.~C. 2014, \aap, 567, A54

\bibitem[{{Rossow} \& {Schiffer}(1999)}]{rossow1999}
{Rossow}, W.~B., \& {Schiffer}, R.~A. 1999, Bull. Americ. Meteor. Soc., 80,
  2261

\bibitem[{{Rothschild} \& {Mancinelli}(2001)}]{rothschild2001}
{Rothschild}, L.~J., \& {Mancinelli}, R.~L. 2001, \nat, 409, 1092

\bibitem[{{S{\'a}ndor} {et~al.}(2007){S{\'a}ndor}, {S{\"u}li}, {{\'E}rdi},
  {Pilat-Lohinger}, \& {Dvorak}}]{sandor2007}
{S{\'a}ndor}, Z., {S{\"u}li}, {\'A}., {{\'E}rdi}, B., {Pilat-Lohinger}, E., \&
  {Dvorak}, R. 2007, \mnras, 375, 1495

\bibitem[{{Schulze-Makuch} {et~al.}(2011){Schulze-Makuch}, {M{\'e}ndez},
  {Fair{\'e}n}, {von Paris}, {Turse}, {Boyer}, {Davila}, {Ant{\'o}nio},
  {Catling}, \& {Irwin}}]{schulzemakuch2011}
{Schulze-Makuch}, D., {M{\'e}ndez}, A., {Fair{\'e}n}, A.~G., {et~al.} 2011,
  \astrb, 11, 1041

\bibitem[{{Selsis} {et~al.}(2007{\natexlab{a}}){Selsis}, {Kasting}, {Levrard},
  {Paillet}, {Ribas}, \& {Delfosse}}]{Selsis2007}
{Selsis}, F., {Kasting}, J.~F., {Levrard}, B., {et~al.} 2007{\natexlab{a}},
  \aap, 476, 1373

\bibitem[{{Selsis} {et~al.}(2007{\natexlab{b}}){Selsis}, {Chazelas},
  {Bord{\'e}}, {Ollivier}, {Brachet}, {Decaudin}, {Bouchy}, {Ehrenreich},
  {Grie{\ss}meier}, {Lammer}, {Sotin}, {Grasset}, {Moutou}, {Barge}, {Deleuil},
  {Mawet}, {Despois}, {Kasting}, \& {L{\'e}ger}}]{selsis07b}
{Selsis}, F., {Chazelas}, B., {Bord{\'e}}, P., {et~al.} 2007{\natexlab{b}},
  Icarus, 191, 453

\bibitem[{{Spiegel} {et~al.}(2008){Spiegel}, {Menou}, \&
  {Scharf}}]{spiegel2008}
{Spiegel}, D.~S., {Menou}, K., \& {Scharf}, C.~A. 2008, \apj, 681, 1609

\bibitem[{{Spiegel} {et~al.}(2009){Spiegel}, {Menou}, \&
  {Scharf}}]{spiegel2009}
---. 2009, \apj, 691, 596

\bibitem[{{Spiegel} {et~al.}(2010){Spiegel}, {Raymond}, {Dressing}, {Scharf},
  \& {Mitchell}}]{Spiegel.etal:2010}
{Spiegel}, D.~S., {Raymond}, S.~N., {Dressing}, C.~D., {Scharf}, C.~A., \&
  {Mitchell}, J.~L. 2010, \apj, 721, 1308

\bibitem[{{Stevenson}(1999)}]{stevenson1999}
{Stevenson}, D.~J. 1999, \nat, 400, 32

\bibitem[{{Terquem} \& {Papaloizou}(2007{\natexlab{a}})}]{Terquem.etal:2007}
{Terquem}, C., \& {Papaloizou}, J.~C.~B. 2007{\natexlab{a}}, \apj, 654, 1110

\bibitem[{{Tremaine} \& {Dong}(2012)}]{Tremaine2012}
{Tremaine}, S., \& {Dong}, S. 2012, \aj, 143, 94

\bibitem[{{Turekian} \& {Clark}(1975)}]{turekian1975}
{Turekian}, K.~K., \& {Clark}, Jr., S.~P. 1975, \jas, 32, 1257

\bibitem[{{Udry} {et~al.}(2007){Udry}, {Bonfils}, {Delfosse}, {Forveille},
  {Mayor}, {Perrier}, {Bouchy}, {Lovis}, {Pepe}, {Queloz}, \&
  {Bertaux}}]{udry07}
{Udry}, S., {Bonfils}, X., {Delfosse}, X., {et~al.} 2007, \aap, 469, L43

\bibitem[{{Valencia} {et~al.}(2006){Valencia}, {O'Connell}, \&
  {Sasselov}}]{Valencia.etal:2006}
{Valencia}, D., {O'Connell}, R.~J., \& {Sasselov}, D. 2006, Icarus, 181, 545

\bibitem[{{von Paris} {et~al.}(2013){von Paris}, {Grenfell}, {Rauer}, \&
  {Stock}}]{vparis2013marsn2}
{von Paris}, P., {Grenfell}, J.~L., {Rauer}, H., \& {Stock}, J.~W. 2013,
  \planss, 82, 149

\bibitem[{{von Paris} {et~al.}(2010){von Paris}, {Gebauer}, {Godolt},
  {Grenfell}, {Hedelt}, {Kitzmann}, {Patzer}, {Rauer}, \&
  {Stracke}}]{vparis2010gliese}
{von Paris}, P., {Gebauer}, S., {Godolt}, M., {et~al.} 2010, \aa, 522, A23

\bibitem[{{Walker} {et~al.}(1981){Walker}, {Hays}, \& {Kasting}}]{walker1981}
{Walker}, J.~C.~G., {Hays}, P.~B., \& {Kasting}, J.~F. 1981, \jgr, 86, 9776

\bibitem[{{Walsh} {et~al.}(2011){Walsh}, {Morbidelli}, {Raymond}, {O'Brien}, \&
  {Mandell}}]{walsh11}
{Walsh}, K.~J., {Morbidelli}, A., {Raymond}, S.~N., {O'Brien}, D.~P., \&
  {Mandell}, A.~M. 2011, \nat, 475, 206

\bibitem[{{Walsh} {et~al.}(2012){Walsh}, {Morbidelli}, {Raymond}, {O'Brien}, \&
  {Mandell}}]{walsh12}
---. 2012, Meteoritics and Planetary Science, 47, 1941

\bibitem[{{Ward}(1997)}]{ward97}
{Ward}, W.~R. 1997, Icarus, 126, 261

\bibitem[{{Weidenschilling}(1977)}]{weidenschilling77}
{Weidenschilling}, S.~J. 1977, \apss, 51, 153

\bibitem[{{Weiss} \& {Marcy}(2014)}]{Weiss2014}
{Weiss}, L.~M., \& {Marcy}, G.~W. 2014, \apjl, 783, L6

\bibitem[{{Weiss} {et~al.}(2013{\natexlab{a}}){Weiss}, {Marcy}, {Rowe},
  {Howard}, {Isaacson}, {Fortney}, {Miller}, {Demory}, {Fischer}, {Adams},
  {Dupree}, {Howell}, {Kolbl}, {Johnson}, {Horch}, {Everett}, {Fabrycky}, \&
  {Seager}}]{Weiss2013}
{Weiss}, L.~M., {Marcy}, G.~W., {Rowe}, J.~F., {et~al.} 2013{\natexlab{a}},
  \apj, 768, 14

\bibitem[{{Williams} {et~al.}(1997{\natexlab{a}}){Williams}, {Kasting}, \&
  {Wade}}]{williams1997moons}
{Williams}, D.~M., {Kasting}, J.~F., \& {Wade}, R.~A. 1997{\natexlab{a}}, \nat,
  385, 234
  
\bibitem[{{Williams} \& {Cieza}(2011)}]{Williams.etal:2011}
{Williams}, J.~P., \& {Cieza}, L.~A. 2011, \araa, 49, 67

\bibitem[{{Wordsworth}(2012)}]{wordsworth2012}
{Wordsworth}, R. 2012, \icarus, 219, 267

\bibitem[{{Wordsworth} {et~al.}(2010){Wordsworth}, {Forget}, {Selsis},
  {Madeleine}, {Millour}, \& {Eymet}}]{wordsworth2010}
{Wordsworth}, R., {Forget}, F., {Selsis}, F., {et~al.} 2010, \aa, 522, A22

\bibitem[{{Wordsworth} {et~al.}(2011){Wordsworth}, {Forget}, {Selsis},
  {Millour}, {Charnay}, \& {Madeleine}}]{wordsworth2011}
{Wordsworth}, R.~D., {Forget}, F., {Selsis}, F., {et~al.} 2011, \apjl, 733, L48

\bibitem[{{Yang} {et~al.}(2014){Yang}, {Bou{\'e}}, {Fabrycky}, \&
  {Abbot}}]{yang2014}
{Yang}, J., {Bou{\'e}}, G., {Fabrycky}, D.~C., \& {Abbot}, D.~S. 2014, \apjl,
  787, L2

\bibitem[{{Yang} {et~al.}(2013){Yang}, {Cowan}, \& {Abbot}}]{yang2013}
{Yang}, J., {Cowan}, N.~B., \& {Abbot}, D.~S. 2013, \apjl, 771, L45

\end{thebibliography}

\end{document}